\documentclass[envcountsect,11pt,letterpaper]{llncs}
\usepackage{soul}
\usepackage[utf8]{inputenc}
\usepackage[table]{xcolor}
 \usepackage{cancel} 
\usepackage{float}
\usepackage[margin=1in]{geometry}
\usepackage{graphicx}
\usepackage{xcolor}
\usepackage{amsmath}
\usepackage{amssymb}
\usepackage[sort,nocompress]{cite}
\usepackage[ruled,vlined,lined,boxed,commentsnumbered]{algorithm2e}
\usepackage{todonotes}
\usepackage{verbatim}
\usepackage{subcaption}
\usepackage[normalem]{ulem}
\usepackage{array}
\usepackage{algorithmic}
\usepackage{mathrsfs}
 \usepackage{multirow}
 \usepackage{blindtext}
\usepackage{hyperref}
\usepackage{graphicx}
\usepackage{todonotes}

\usepackage{pgf}
\usepackage{tikz}
\usepackage{tikz-cd}
\tikzcdset{scale cd/.style={every label/.append style={scale=#1},
    cells={nodes={scale=#1}}}}

\usetikzlibrary{arrows,automata,positioning}
\SetKwComment{Comment}{$\triangleright$\ }{}

\title{Graphs can be succinctly indexed for pattern matching in $ O(|E|^2 + |V|^{5 / 2}) $ time}

\author{
Nicola Cotumaccio
} 

\institute{
Gran Sasso Science Institute (GSSI), L'Aquila,  Italy. Email: \email{nicola.cotumaccio@gssi.it}}

\date{\today}

\begin{document}

\maketitle
\thispagestyle{empty}

\begin{abstract}

Pattern matching is a pervasive problem in computer science. In several applications (notably, bioinformatics), it is crucial to perform pattern matching on edge-labeled graphs: given a string, decide whether the string can be read on the graph. In this paper we consider the problem of indexing (preprocessing) a graph for pattern matching by building a \emph{succinct} data structure, that is, a data structure that uses a number of bits close to the lower bounds from information theory, while allowing efficient queries. In [TCS 2017] Gagie et al. showed how to index a class of graphs - the so-called \emph{Wheeler graphs} - by totally ordering the nodes of the graph (building a \emph{Wheeler order}) and using some techniques based on the Burrows-Wheeler transform. However, the class of Wheeler graphs is rather small: for example, a unary language is recognized by some Wheeler automaton (that is, an automaton whose underlying graph is Wheeler) if and only if the language is finite or cofinite. In [SODA 2021] these ideas where extended to arbitrary node-labeled graphs, and Wheeler orders were generalized to \emph{co-lex orders}, which are \emph{partial} orders: it was showed that a node-labeled graphs $ G = (V, E) $ can be succinctly indexed by means of a data structure of $|E|(\lceil \log|\Sigma|\rceil + \lceil\log p\rceil + 2)\cdot (1+o(1)) + |V|\cdot (1+o(1))$ bits which supports pattern matching in $O(|P| \cdot p^2 \cdot \log(p\cdot |\Sigma|))$ time, where $ p $ is the minimum width of a co-lex order on $ G $, $ P $ is the pattern and $ \Sigma $ is the alphabet. However, determining $ p $ is NP-hard and building the data structure is also hard. Intuitively, a partial order imposes antisymmetry and transitivity constraints, which cause the hardness of all natural decision problems connected to Wheeler and co-lex orders. In this paper, we change perspective switching from (partial) orders to \emph{arbitrary} relations, so defining \emph{co-lex relations}. We work in the more general setting of \emph{edge}-labeled graphs $ G = (V, E) $, and for the first time we provide a \emph{succinct} index for \emph{arbitrary} graphs that can be built \emph{in polynomial time}, which requires less space and answers queries more efficiently than the one in [SODA 2021]. We show that, given an edge-labeled graph $ G = (V, E) $, there exists a data structure of $|E /_{\le_G}|(\lceil \log|\Sigma|\rceil + \lceil\log q\rceil + 2)\cdot (1+o(1)) + |V /_{\le_G}|\cdot (1+o(1))$ bits which supports pattern matching on $ G $ in $O(|P| \cdot q^2 \cdot \log(q\cdot |\Sigma|))$ time, where  $ G /_{\le_G} = (V /_{\le_G}, E /_{\le_G}) $ is a quotient graph obtained by collapsing some nodes in $ G $ (so $ |V /_{\le_G}| \le |V| $ and $ |E /_{\le_G}| \le |E| $) and $ q $ is the width of the \emph{maximum} co-lex relation on $ G $. The bounds achieved in this paper look similar to the ones in [SODA 2021], but, in fact, there are several sources of improvement:
\begin{enumerate}
    \item Most importantly, $ q $ can be determined in $ O(|E|^2) $ time, and our data structure can be built in $ O(|E|^2 + |V /_{\le_G}|^{5 / 2}) $ time (while determining $ p $ and building the data structure in [SODA 2021] are hard problems).
    \item It always holds $ q \le p $, and $ q $ can be arbitrarily smaller than $ p $ (that is, for every integer $ n $ there exists a graph such that $ p = n $ and $ q = 1 $).
    \item Our bound only depends on the size of $ G /_{\le_G} $ and it is independent of the size of $ G $. In other words, the quotient graph \emph{$ G /_{\le_G} $ eliminates the unnecessary redundancy in $ G $ from a pattern matching perspective}.
\end{enumerate}
Our results have relevant applications in automata theory. First, we can build a succinct data structure to decide whether a string is accepted by a given automaton. Second, starting from an automaton $ \mathcal{A} $, one can define a relation $ \preceq_\mathcal{A} $ and a quotient automaton that capture the \emph{nondeterminism} of $ \mathcal{A} $, improving the results in [SODA 2021].
\end{abstract}

\newpage

\setcounter{page}{1}

\section{Introduction}

Pattern matching is a pervasive problem in computer science: given a pattern and some data, decide whether the pattern matches the data. In this paper, we will consider graph pattern matching.

\paragraph*{\textbf{Graph pattern matching}} Let $ \Sigma $ be an alphabet, and let $ G = (V, E) $ be an edge-labeled graph. The \emph{pattern-matching problem} is the following: given a pattern $ P \in \Sigma^* $, decide whether $ \alpha $ can be read on $ G $ by following edges whose labels, when concatenated, yield $ P $.

\paragraph*{}

The problem of matching patterns on graphs arises in a number of fields. In bioinformatics, the pan-genome is a labeled graph capturing the genetic variation within a species \cite{baier, siren}. Pattern matching on labeled graphs is also natural in graph databases \cite{williams, angles}.

In this paper we consider the problem of building an \emph{index} for pattern matching. In other words, we aim to preprocess a given graph in such a way that we can quicker answer multiple pattern matching queries. At the same time, we want to employ a \emph{succinct} data structure \cite{navarro_2016}, that is, a data structure that requires a number of bits close to the lower bounds from information theory, while allowing efficient queries. Several succinct data structures used for indexing rely on some variant of the Burrows-Wheeler Transform (BWT) \cite{burrows}.

The problem of indexing \emph{node}-labeled graphs for pattern matching has been extensively studied in the last years. When searching for a match on node-labeled graphs, one still follows edges, and a string is read by concatenating the labels on the nodes. The problem of pattern matching on edge-labeled graphs is more general, because a node-labeled graph can be thought of as an edge labeled graph where all edges entering the same node have the same label.

On the one hand, in \cite{equi2020graphs} Equi et al. showed that there exists no algorithm indexing a node-labeled graph $ G = (V, E) $ in polynomial time in such a way that a pattern $ P \in \Sigma^*$ can be checked for pattern matching  in $ O(|E|^\delta |P|^\beta) $ time, where $ \delta < 1 $ or  $ \beta < 1$, unless the Orthogonal Vector Hypothesis (OVH) is false. On the other hand, in \cite{GAGIE201767} Gagie et al. introduced a class of graphs - the so-called \emph{Wheeler graphs} - which can be succinctly stored while allowing pattern matching in $ O(|P| \log |\Sigma|) $ time. Wheeler graphs generalize a number of previous approaches based on \emph{sorting} the nodes of the graph, and then applying BWT-like techniques to efficiently support pattern matching. Such graphs are endowed with a total order (a \emph{Wheeler order}) that satisfies \emph{path coherence}: if one starts from an interval of nodes and follow all edges labeled with a letter $ a $, then one still ends up in an interval of nodes. Note that the bound $ O(|P| \log |\Sigma|) $ does not break the bound $ O(|E|^\delta |P|^\beta) $ because most graphs are not Wheeler graphs. For example, it can be showed that a unary language is recognized by some Wheeler automaton (that is, an automaton whose underlying graph is Wheeler) if and only if the language is finite or cofinite \cite{ alanko2020wheeler}. Another limitation of Wheeler graphs is that all natural problems connected with the property of being Wheeler are hard: in, particular, deciding whether a node-labeled graph is Wheeler is an NP-complete problem \cite{gibney2019ESA}.

In \cite{cotumaccio2021indexing} the indexing techniques for Wheeler graphs were extended to arbitrary node-labeled graphs. The main idea is to consider orders on the set of nodes that are allowed to be \emph{partial}, the so-called \emph{co-lex orders}. In general, if $ (V, \le) $ is a partial order, we can consider a partition $ \{V_i \}_{i = 1}^p $ of $ V $ such that, for every $ i $,  every pair of elements in $ V_i $ are $ \le $-comparable. The minimum size of such a partition is the \emph{width} of the partial order. It can be showed that a co-lex order of width $ p $ can be used to succinctly index the graph in a such a way that pattern matching queries can be solved in $O(|P| \cdot p^2 \cdot \log(p\cdot |\Sigma|)) $ time. In particular, a co-lex order is a Wheeler order if and only if $ p = 1 $, and in this case we retrieve the bound $ O(|P| \log |\Sigma|) $. While most node-labeled graph do not admit a Wheeler order, every node-labeled graph admits a co-lex order. Both the bound $O(|P| \cdot p^2 \cdot \log(p\cdot |\Sigma|)) $ and the (succinct) number of bits required to index the graph are proportional to $ p $, so one should determine a co-lex order of width as small as possible. However, co-lex orders inherit the hardness of the problems connected to Wheeler orders: determining the minimum width of a co-lex order on a graph is NP-hard (we mentioned that the simpler problem of determining whether a graph is Wheeler is already NP-complete). This implies that the problem of indexing a graph with the best (i.e., minimum-width) co-lex order is hard. In addition, no approximation algorithm for computing such a minimum width (and the corresponding co-lex order) is currently known. 

\section{Our contribution}\label{sec:contributions}

Wheeler orders and co-lex orders allow efficient pattern matching because they ensure (some variant of) path coherence. Since sorting the nodes of graphs leads to place restrictions - such as Wheelerness - being NP-hard to check, it is natural to wonder whether it is in fact necessary to rely on sorting. Defining a total order implies ensuring antisymmetry and transitivity, and these properties are logically hard to express. For example, in \cite{alanko2020regular} it was showed that on a special class of graphs (2-NFAs) the problem of deciding whether a given graph is Wheeler can be solved in polynomial time, and the main idea is to reduce the problem to 2-SAT by defining clauses expressing the property of being Wheeler. The reason why this method does not work for arbitrary graphs is that one should in particular define a clause for expressing transitivity, and transitivity requires 3-SAT clauses on general graphs. Similarly, antisymmetry is not a necessary constraint from a pattern matching perspective, because we will show that if two nodes are comparable in both directions, then they can be essentially thought of as a unique node.

To sum up, in this work we change perspective switching from (partial) orders to \emph{arbitrary} relations, so defining \emph{co-lex relations}. By removing antisymmetry and transitivity, we show that the notion of path coherence still makes perfect sense, and the algebraic structure behind pattern matching becomes cleaner. Indeed, we show that every graph $ G $ admits a \emph{maximum} co-lex relation, that is, a co-lex relation $ R$  such that \emph{every} co-lex relation on $ G $ is a restriction of $ R $ (while in general a graph does \emph{not} admit a maximum co-lex \emph{order}). In particular, the width of the maximum co-lex relation is automatically the minimum width of a co-lex relation on the graph. Moreover (1) the maximum co-lex relation can be computed in $ O(|E|^2) $ time and (2) it is always transitive. While transitivity is not conceptually relevant for path coherence, from an algorithmic perspective it is helpful for indexing. Moreover, we show that all nodes being comparable in both directions can be compressed into a single node. More precisely, we show that starting from a graph $ G $ one can always build a \emph{quotient} graph $ G /_{\le_G} $ that captures exactly the same information for pattern matching: one can always answer a query on $ G $ by answering the same query on $ G /_{\le_G} $. This approach is successful because the graph $ G /_{\le_G} $ is topologically simpler than the original graph $ G $: if a node in $ G /_{\le_G} $ has been obtained by collapsing two or more nodes of the original graph, than such a node can have at most one ingoing edge in the quotient graph. Moreover, $ G /_{\le_G} $ always admits a maximum co-lex \emph{order} (while a general graph does not admit a maximum co-lex order, as stated above), which is naturally induced by the maximum co-lex relation on $ G $. Since $ G /_{\le_G} $ admits the maximum co-lex order and, crucially, it can be built \emph{in polynomial time}, we can index $ G $ by simply indexing $ G /_{\le_G} $ using the techniques from \cite{cotumaccio2021indexing}.

Let us state our quantitative results. In \cite{cotumaccio2021indexing} it was showed that a node-labeled graphs $ G = (V, E) $ can be succinctly indexed by means of a data structure of $|E|(\lceil \log|\Sigma|\rceil + \lceil\log p\rceil + 2)\cdot (1+o(1)) + |V|\cdot (1+o(1))$ bits which supports pattern matching in $O(|P| \cdot p^2 \cdot \log(p\cdot |\Sigma|))$ time, where $ p $ is the minimum width of a co-lex order on $ G $, $ P $ is the pattern and $ \Sigma $ is the alphabet. Determining $ p $ is NP-hard and building the data structure is also hard. In this paper, we work in the more general setting of \emph{edge}-labeled graphs $ G = (V, E) $, and for the first time we provide a \emph{succinct} index for \emph{arbitrary} graphs that can be built \emph{in polynomial time} (while even only determining if a graph is Wheeler is NP-hard), which requires less space and answers queries more efficiently than the one in \cite{cotumaccio2021indexing}. We show that, given an edge-labeled graph $ G = (V, E) $, there exists a data structure of $|E /_{\le_G}|(\lceil \log|\Sigma|\rceil + \lceil\log q\rceil + 2)\cdot (1+o(1)) + |V /_{\le_G}|\cdot (1+o(1))$ bits which supports pattern matching on $ G $ in $O(|P| \cdot q^2 \cdot \log(q\cdot |\Sigma|))$ time, where  $ G /_{\le_G} = (V /_{\le_G}, E /_{\le_G}) $ is the quotient graph (so $ |V /_{\le_G}| \le |V| $ and $ |E /_{\le_G}| \le |E| $) and $ q $ is the width of the maximum co-lex relation on $ G $. The bounds achieved in this paper look similar to the ones in \cite{cotumaccio2021indexing}, but, in fact, there are several sources of improvement:
\begin{enumerate}
    \item Most importantly, $ q $ can be determined in $ O(|E|^2) $ time, and our data structure can be built in $ O(|E|^2 + |V /_{\le_G}|^{5 / 2}) $ time (while determining $ p $ and building the data structure in \cite{cotumaccio2021indexing} are hard problems).
    \item It always holds $ q \le p $, and $ q $ can be arbitrarily smaller than $ p $ (that is, for every integer $ n $ there exists a graph such that $ p = n $ and $ q = 1 $).
    \item Our bound only depends on the size of $ G /_{\le_G} $ and it is independent of the size of $ G $. In other words, \emph{$ G /_{\le_G} $ eliminates the unnecessary redundancy for pattern matching}.
\end{enumerate}

We point out that in this paper we do not propose new data structures, but we show that co-lex relations allow to define a quotient graph on which the data structure introduced in \cite{cotumaccio2021indexing} can be built in polynomial time and performs better.

Next, we show that our indexing techniques are flexible enough to solve more general problems, such as deciding whether a string is accepted by a given automaton by means of a succinct data structure. In particular, we define a quotient automaton that recognizes the same language. More generally, we prove that the width of a natural relation on the states of a NFA captures the degree of nondeterminism of the automaton. In \cite{cotumaccio2021indexing} it was showed that the powerset DFA that determinizes a given NFA has a number of states which is exponential only in the minimum width $ p $ of a co-lex order on the NFA. In this paper, we prove that, in fact, there is a deeper and simpler quantity capturing the blow-up in the powerset construction, which is simply the width of a relation. Additionally, one can canonically build a quotient automaton that captures the nondeterminism of a given automaton - that is, the given automaton and its quotient have the same powerset automaton.

\section{Notation}

Let $ \Sigma $ be an alphabet, and let $ \preceq $ be a fixed, total order on $ \Sigma $. We denote by $ G = (V, E) $ an (edge-labeled) graph, where $ V $ is the set of nodes, and $ E \subseteq V \times V \times \Sigma $ is the set of labeled edges. In this papers, all graphs are finite.

If $ V $ is a set, a \emph{(binary) relation} $ R $ on $ V $ is a subset of $ V \times V $. We say that $ u, v \in V $ are \emph{$ R $-comparable} if $ (u, v) \in R \lor (v, u) \in R $ (note that $ (u, v) \in R $ and $ (v, u) \in R $ may be \emph{both} true). We denote by $ Trans (R) $ the transitive closure of $ R $. If $ R $ and $ R' $ are binary relations on $ V $, we say that $ R $ \emph{refines} $ R' $ if $ (u, v) \in R' \Rightarrow (u, v) \in R $. If $ R $ is a binary relation on $ V $ and $ U \subseteq V $, we say that $ U $ is \emph{$ R $-convex} if:
\begin{equation*}
    (\forall u, v, z \in V)((u, z \in U \land (u, v) \in R \land (v, z) \in R) \implies v \in U).
\end{equation*}

A \emph{preorder} $ \le $ on $ V $ is a binary relation being reflexive and transitive. We write $ u < v $ if $ u \le v $ and $ u \not = v $. Moreover, the preorder $ \le $ is a \emph{partial order} if it antisymmetric, and it is a \emph{total order} if it is a partial order and every pair of elements are $ \le $-comparable.

We introduce some notation typical of partial order, and we naturally extend it to preorders. Let $ (V, \le) $ be a preorder. A set $ V' \subseteq V $ is a \emph{$ \le $-chain} if every $ u, v \in V $ are $ \le $-comparable. A set $ V' \subseteq V $ is a \emph{$ \le $-antichain} if every distinct $ u, v \in V $ are not $ \le $-comparable. A partition $ \{V_i \}_{i = 1}^p $ of $ V $ is a \emph{$ \le $-chain partition} if every $ V_i $ is a $ \le $-chain. The \emph{width} of $ (V, \le) $ is the minimum size of a $ \le $-chain partition. Note that if $ \le $ and $ \le' $ are preorders on $ V $, and $ \le $ refines $ \le' $, then the width of $ \le $ is smaller than or equal to the width of $ \le' $ (because every $ \le' $-chain partition is also a $ \le $-chain partition). If $ (V, \le) $ is a partial order, then \emph{Dilworth's theorem} \cite{dilworth2009decomposition} states that the width of $ (V, \le) $ is equal to the maximum size of a $ \le $-antichain.

Let us recall a standard method for obtaining a partially-ordered quotient set from a preorder. Let $ (V, \le) $ be a preorder. For every $ u, v \in V $, let $ u \sim_\le v $ if and only if $ (u \le v) \land (v \le u) $. It is immediate to check that $ \sim_\le $ is an equivalence relation. Now, let $ [v]_\le $ be the quotient class of $ v $, and consider the quotient set $ V /_\le = \{[v]_\le | v \in V \} $. Define $ \le^{\sim} $ on $ V /_\le $ by letting $ [u]_\le \le^\sim [v]_\le $ if and only if $ u \le v $. The definition of $ \sim_\le $ implies that $ \le^\sim $ is well-defined (that is, the definition does not depend on the choice of representatives), because if $ u \sim_\le u' $, $ v \sim_\le v' $ and $ u \le v $, then $ u' \le u \le v \le v' $. Moreover $ (V /_\sim , \le^\sim) $ is a \emph{partial} order. Indeed, if $ [u]_\le \le^\sim [v]_\le $ and $ [v]_\le \le^\sim [u]_\le $, then $ u \le v $ and $ v \le u $, so $ [u]_\le = [v]_\le $.

We denote by $ \mathcal{A} = (Q, E, s, F) $ a \emph{non-deterministic finite automaton} (NFA), where $ Q $ is the set of states, $ E \subseteq Q \times Q \times \Sigma $ is the set of edges, $ s \in Q $ is the initial state and $ F \subseteq Q $ is the set of final states. A \emph{deterministic finite automaton} (DFA) is an NFA such that for every $ u \in Q $ and for every $ a \in \Sigma $ there exists at most one $ v \in Q $ such that $ (u, v, a) \in E $. As customary (for example in DFA minimization), we assume that all states are reachable from the initial state and each state is either final, or it allows to reach a final state (states violating these assumptions can be removed by means of a graph visit without changing the recognized language).  Let $ \mathcal{L(A)} $ be the regular language recognized by $ \mathcal{A} $. Let $ Pref (\mathcal{L(A))} $ the set of all prefixes of some word in $ \mathcal{L(A)} $, and for $ u \in Q $ let $ I_u $ be the set of all strings that can be read from $ s $ to $ u $. Our assumptions imply that $ \{I_u | u \in Q \} $ is a \emph{cover} of $ Pref (\mathcal{L(A)}) $ (if $ \mathcal{A} $ is a DFA, then $ \{I_u | u \in Q \} $ is a \emph{partition} of $ Pref (\mathcal{L(A)}) $). Moreover, for $ \alpha \in Pref (\mathcal{L(A)}) $ denote by $ I_\alpha $ the set of all states $ u \in Q $ such that $ \alpha \in I_u $.

Recall that, startig from an NFA $ \mathcal{A} $, the \emph{powerset construction} algorithm builds a DFA $ \mathcal{A^*} = (Q^*, E^*, s^*, F^*) $ such that $ \mathcal{L(\mathcal{A})} = \mathcal{L(\mathcal{A^*})} $ defined as follows:
(i) $ Q^* = \{ I_{\alpha}\ |\ \alpha\in Pref (\mathcal{L(\mathcal{A})})\}$,
(ii) $ E^* = \{(I_\alpha, I_{\alpha a}, a) \in Q^* \times Q^* \times \Sigma | \alpha \in \Sigma^*, a \in \Sigma, \alpha a \in Pref (\mathcal{L(\mathcal{A})}) \} $, (iii) $ s^* = I_\epsilon = \{s \} $, (iv) $ F^* = \{I_\alpha\ |\ \alpha \in \mathcal{L(\mathcal{A})} \} $. Moreover, if for $ q^* \in Q^* $ we let $ I^*_{q^*} $ be the set of all strings in $ Pref (\mathcal{L(A)}) $ that can be read from $ s^* $ to $ u^* $ on $ \mathcal{A^*} $, then for every $ \alpha \in Pref (\mathcal{L(A)}) $:
\begin{equation}\label{eq:powersetstates}
    I^*_{I_\alpha} = \{\alpha' \in Pref (\mathcal{L(A)}) | I_{\alpha'} = I_\alpha \}.
\end{equation}

\section{Definitions and first results}

Let $ G = (V, E) $ be a graph. Let $ \# \not \in \Sigma $ be a special symbol, and assume $ \# \prec a $ for all $ a \in \Sigma $. For $ v \in V $ define:
\begin{equation*}
    \lambda (v) = 
    \begin{cases}
        \{a \in \Sigma^* | \text{ $ (u, v, a) \in E $ for some $ u \in V $} \} & \text{ if $ v $ has incoming edges} \\
        \{\# \} & \text{ if $ v $ does not have incoming edges}.
    \end{cases}
\end{equation*}

In a Wheeler order, all nodes without incoming edges must come before all remaining nodes \cite{GAGIE201767}. Intuitively, we let $ \# \prec a $ for all $ a \in \Sigma $ to ensure a similar properties for arbitrary relations.

If $ u, v \in Q $, define:
\begin{equation*}
    \lambda (u) \; \angle \; \lambda (v) \iff (\forall a \in \lambda (u))(\forall b \in \lambda (v))(a \preceq b).
\end{equation*}

\begin{remark}\label{rem:propertiesangletry}
Notice that (1) for every $ v \in V $ it holds $ \lambda (v) \not = \emptyset $; (2) if $ \lambda (u) \; \angle \; \lambda (v) $ and $ \lambda (v) \; \angle \; \lambda (z) $, then $ \lambda (u) \; \angle \; \lambda (z) $; (3) if $ \lambda (u) \; \angle \; \lambda (v) $ and $ \lambda (v) \; \angle \; \lambda (u) $, then $ \lambda (u) = \lambda (v) $ and $ |\lambda (u)| = |\lambda (v)| = 1 $. (4) If $ \lambda (u) = \lambda (v) $, then $ \lambda (u) \; \angle \; \lambda (v) $ (and $ \lambda (v) \; \angle \; \lambda (u) $) if and only if $ |\lambda (u)| = |\lambda (v)| = 1 $.
\end{remark}

We can now give our main definition, which generalizes the definition of co-lex order given in \cite{cotumaccio2021indexing} (the term "co-lex order" refers to the co-lexicographic ordering of strings induced by a co-lex order, see Section \ref{sec:automata}).

\begin{definition}
Let $ G = (V, E) $ be a graph. A \emph{co-lex relation} on $ G $ is a reflexive relation $ R \subseteq V \times V $ that satisfies the following two axioms:
\begin{enumerate}
    \item (Axiom 1) For every $ u, v \in Q $ such that $ u \not = v $, if $ (u, v) \in R $, then $ \lambda (u) \; \angle \; \lambda (v) $;
    \item (Axiom 2) For every $ (u', u, a), (v', v, a) \in E $ such that $ u \not = v $, if $ (u, v) \in R $, then $ (u', v') \in R $.
\end{enumerate}
A \emph{co-lex preorder} is a co-lex relation that is also a preorder. A \emph{co-lex order} is a co-lex relation that is also a partial order.
\end{definition}

\begin{remark}\label{rem:initialproperty}
(1) Let $ u \in V $ be a state with no incoming edges, and let $ v \in V $ a state with incoming edges. From Axiom 1, it follows $ (v, u) \not \in R $. (2) If for distinct $ u, v \in Q $ it holds $ (u, v) \in R $ and $ (v, u) \in R $, then by Axiom 1 and Remark \ref{rem:propertiesangletry} we conclude $ \lambda (u) = \lambda (v) $ and $ |\lambda (u)| = |\lambda (v)| = 1 $.
\end{remark}

\begin{remark}\label{rem:existencecolexrelation}
Every graph $ G = (V, E) $ admits a co-lex relation. For example, $ \{(v, v) | v \in V \} $ and $ \{ (u, v) \in V \times V | (\forall a \in \lambda (u))(\forall b \in \lambda (v))(a \prec b) \} \cup \{(v, v) | v \in V \} $  are co-lex relations on $ G $. 
\end{remark}

The property that allows to index a Wheeler graph for pattern matching is \emph{path coherence}: starting from an interval of nodes and reading a string one still ends up in an interval of nodes \cite{GAGIE201767}. This property was generalized to co-lex orders \cite{cotumaccio2021indexing}, and we now generalize it to arbitrary co-lex relations.

\begin{lemma}[Path coherence]\label{lem:pathcoherencecolexrelations}
Let $ G = (V, E) $ be a graph, and let $ R $ be a co-lex relation on $ G $. Let $ \alpha \in \Sigma^*$, and let $ U \subseteq V $ be $ R $-convex. Then, the set $ U' $ of all nodes in $ V $ that can be reached from $ U $ by following edges whose labels, when concatenated, yield $ \alpha $, is still $ R $-convex (possibly $ U' $ is empty).
\end{lemma}

\begin{proof}
We proceed by induction on $ | \alpha | $. If $ |\alpha| = 0 $, then $ \alpha = \epsilon $ and we are done. Now assume $ |\alpha| \ge 1 $. We can write $ \alpha = \alpha' a $, with $ \alpha' \in \Sigma^* $, $ a \in \Sigma $. Let $ u, v, z \in Q $ such that $ u, z \in U' $ and $ (u, v), (v, z) \in R $. We must prove that $ v \in U' $. If $ v = u $ or $ v = z $ the conclusion follows, so we can assume $ v \not = u $ and $ v \not = z $. By the inductive hypothesis, the set $ U'' $ of all nodes in $ V $ that can be reached from some state in $ U $ by following edges whose labels, when concatenated, yield $ \alpha' $, is $ R $-convex. In particular, there exist $ u', z' \in U'' $ such that $ (u', u, a) \in E $ and $ (z', z, a) \in E $. Since $ a \in \lambda (u) \cap \lambda (z) $ and $ (u, v), (v, z) \in R $, then $ \lambda (v) = \{a\} $ (otherwise by Axiom 1 we would obtain a contradiction), so there exists $ v' \in V $ such that $ (v', v, a) \in E $. From $ (u, v), (v,z) \in R $ and Axiom 2 we obtain $ (u', v'), (v', z') \in R $; since $ u', z' \in U'' $ and $ U'' $ is $ R $-convex, then $ v' \in U''$, which implies $ v \in U' $. \qed
\end{proof}

We can already observe that switching from co-lex orders to co-lex relations simplifies the algebraic structure. In general, the union of two co-lex orders is not a co-lex order (see Figure \ref{fig:maximumcolexorder}). However, the union of two co-lex relations is always a co-lex relation:

\begin{figure}
	\centering
	     \begin{subfigure}[b]{0.4\textwidth}
         \centering
	\begin{tikzpicture}[shorten >=1pt,node distance=1.6cm,on grid,auto]
	\tikzstyle{every state}=[fill={rgb:black,1;white,10}]
	
	\node[state]   (q_0)                          {$ 0 $};
	\node[state]           (q_1)  [above right of=q_0]    {$ 1 $};
	\node[state]           (q_2)  [below right of=q_0]    {$ 2 $};

	\path[->]
	(q_0) edge node {a}    (q_1)
	(q_0) edge node {a}    (q_2);
	\end{tikzpicture}
     \end{subfigure}
     \hfill
     	     \begin{subfigure}[b]{0.4\textwidth}
         \centering
	\begin{tikzpicture}[shorten >=1pt,node distance=1.6cm,on grid,auto]
	\tikzstyle{every state}=[fill={rgb:black,1;white,10}]
	
	\node[state]           (q_0)     {$ 0 $};
	\node[state]           (q_1)  [right of=q_1]    {$ 1 $};

	\path[->]
	(q_0) edge [bend left = 90]  node {a}    (q_1)
	(q_1) edge [bend left = 90] node {a}    (q_0);
	\end{tikzpicture}
     \end{subfigure}
	\caption{\emph{Left}: Notice that $ \{(0, 0), (1, 1), (2, 2), (1, 2)  \} $ and $ \{(0, 0), (1, 1), (2, 2), (2, 1)  \} $ are co-lex orders, but their union is not a co-lex order (antisymmetry would be violated). In particular, the graph does not admit the maximum co-lex order. \emph{Right}: A graph that admits the maximum co-lex order, which however is distinct from the maximum co-lex relation. Indeed, the maximum co-lex order is $ \{(0, 0), (1, 1) \} $ and the maximum co-lex relation is $ \{(0, 0), (1, 1), (0, 1), (1, 0) \} $.}\label{fig:maximumcolexorder}
\end{figure}
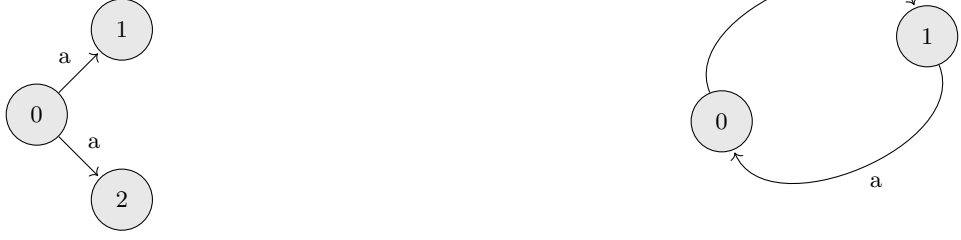

\begin{lemma}\label{lem:union}
Let $ G = (V, E) $ be a graph, and let $ R_1, \dots, R_m $ be co-lex relations on $ G $. Then, $ \cup_{i = 1}^m R_i $ is a co-lex relation on $ G $.
\end{lemma}

\begin{proof}
First, $ \cup_{i = 1}^m R_i $ is reflexive because each $ R_i $ is reflexive. Let us prove Axiom 1. Assume that $ (u, v) \in \cup_{i = 1}^m R_i $, with $ u \not = v $. We must prove that $ \lambda (u) \; \angle \; \lambda (v) $. Notice that it must be $ (u, v) \in R_j $ for some $ j $, so the conclusion follows from Axiom 1 applied to the co-lex relation $ R_j $. Let us prove Axiom 2. Assume that $ (u', u, a), (v', v, a) $ are such that $ u \not = v $ and $ (u, v) \in \cup_{i = 1}^m R_i $. We must prove that $ (u', v') \in \cup_{i = 1}^m R_i $. Notice that it must be $ (u, v) \in R_j $ for some $ j $, so the conclusion follows from Axiom 2 applied to the co-lex relation $ R_j $. \qed
\end{proof}

We now prove that every co-lex relation is refined by a co-lex preorder, namely, its transitive closure.

\begin{lemma}\label{lem:transitiveclosurecolex}
Let $ G = (V, E) $ be a graph, and let $ R $ be a co-lex relation on $ G $. Then, $ Trans(R) $ is a co-lex preorder on $ G $.
\end{lemma}

\begin{proof}
First, $ Trans (R) $ is reflexive because $ R $ is reflexive, and it is a partial order by definition.

Let us prove Axiom 1. Assume that $ (u, v) \in Trans(R) $, with $ u \not = v $. We must prove that $ \lambda(u) \; \angle \; \lambda (v) $. Since $ (u, v) \in Trans(R) $, then there exist $ z_1, \dots, z_r \in Q $ ($ r \ge 0 $) such that $ (u, z_1) \in R $, $ (z_1, z_2) \in R $, $ \dots $, $ (z_r, v) \in R $ and $ u \not = z_1 $, $ z_1 \not = z_2 $, $ \dots $, $ z_r \not = v $. Then, Axiom 1 applied to $ R $ implies $ \lambda (u) \; \angle \; \lambda (z_1) $, $ \lambda (z_1) \; \angle \; \lambda (z_2) $, $ \dots $, $ \lambda (z_r) \; \angle \; \lambda (v) $, so we conclude $ \lambda (u) \; \angle \; \lambda (v) $ by Remark \ref{rem:propertiesangletry}.

Let us prove Axiom 2. Assume that $ (u', u, a), (v', v, a) \in E $ are such that $ u \not = v $ and $ (u, v) \in Trans(R) $. We must prove that $ (u', v') \in Trans (R) $. Since $ (u, v) \in Trans(R) $, then like before there exist $ z_1, \dots, z_r \in Q $ ($ r \ge 0 $) such that $ (u, z_1) \in R $, $ (z_1, z_2) \in R $, $ \dots $, $ (z_r, v) \in R $ and $ u \not = z_1 $, $ z_1 \not = z_2 $, $ \dots $, $ z_r \not = v $, and it must be $ \lambda (u) \; \angle \; \lambda (z_1) $, $ \lambda (z_1) \; \angle \; \lambda (z_2) $, $ \dots $, $ \lambda (z_r) \; \angle \; \lambda (v) $. Since $ a \in \lambda (u) \cap \lambda (v) $, we conclude $ \lambda (z_1) = \dots = \lambda (z_r) = \{a \} $. This implies that there exist $ z'_1, \dots, z'_r \in Q $ such that $ (z'_1, z_1, a) \in E $, $ \dots $, $ (z'_r, z_r, a) \in E $. Then, Axiom 2 applied to $ R $ implies $ (u', z'_1) \in R $, $ (z'_1, z'_2) \in R $, $ \dots $, $ (z'_r, v') \in R $, so we conclude $ (u', v') \in Trans (R) $. \qed
\end{proof}

\begin{definition}
Let $ G = (V, E) $ be a graph. Let $ R $ be a co-lex relation on $ G $. We say that $ R $ is \emph{maximum} if it refines every co-lex relation $ R' $ on $ G $.
\end{definition}

It is clear that if a maximum co-lex relation exists, then it is unique. The following lemma shows that the maximum co-lex relation always exists. This is a crucial distinction between co-lex relations and co-lex orders: in general, the maximum co-lex order - that is, a co-lex order refining every co-lex order - does not exist (see Figure \ref{fig:maximumcolexorder}), and this provides some intuition about why determining the minimum width $ p $ of a co-lex order on a graph is NP-hard.

\begin{lemma}\label{lem:existencemaximumcolexrelation}
Every graph $ G = (V, E) $ admits the maximum co-lex relation (in the following denoted by $ \le_G $). Moreover, $ \le_G $ is a co-lex preorder.
\end{lemma}

\begin{proof}
Let $ \le_G $ be the union of all co-lex relations on $ \mathcal{N} $. Notice that such an union is nonempty by remark \ref{rem:existencecolexrelation} and it is finite because the number of binary relations on $ V $ is finite. Moreover, $ \le_G  $ is a co-lex relation by Lemma \ref{lem:union}, and if for some co-lex relation it holds $ (u, v) \in R $, then by definition $ u \le_G v $, so $ \le_G  $ is the maximum co-lex relation. Finally, $ Trans(\le_G) $ is a co-lex relation by Lemma \ref{lem:transitiveclosurecolex}, so the maximality of $ \le_G $ implies $ Trans(\le_G) = \le_G $, that is, $ \le_G $ is transitive. \qed
\end{proof}

\begin{remark}
Since $ \le_G $ refines every co-lex relation on $ G $, then the width of $ \le_G $ is smaller than or equal to the width of any co-lex relation on $ G $.
\end{remark}

Lemma \ref{lem:existencemaximumcolexrelation} implies that the notion of maximum co-lex preorder (a co-lex preorder refining every co-lex preorder) is pointless, because the maximum co-lex preorder always exists and it is always equal to the maximum co-lex relation. If the maximum co-lex relation is also antisymmetric, then it also the maximum co-lex order; however in general the maximum co-lex order does not exist, or if it exists it can be distinct from the maximum co-lex relation (and in this case the maximum co-lex relation is a strict refinement of the maximum co-lex order), see Figure \ref{fig:maximumcolexorder}.

We now show that the maximum co-lex relation can be computed in $ O(|E|^2) $ time. To this end, we need the characterization in Lemma \ref{lem:precedesmaximum}. Since the maximum co-lex relation is transitive, when indexing a graph we can assume that we use a co-lex preorder.

\begin{definition}\label{def:pairprecedes}
Let $ G = (V, E) $ be a graph, and let $ (u', v'), (u, v) \in V \times V $ be pairs of distinct nodes. We say that $ (u', v') $ \emph{precedes} $ (u, v) $ if there exist $ u_1, \dots, u_r, v_1, \dots, v_r \in V $ ($ r \ge 1 $) and $ a_1, \dots, a_{r - 1} \in \Sigma $ such that:
\begin{enumerate}
    \item $ u_1 = u' $ and $ v_1 = v' $;
    \item $ u_r = u $ and $ v_r = v $;
    \item $ u_i \not = v_i $ for $ i = 1, \dots, r $;
    \item $ (u_i, u_{i + 1}, a_i), (v_i, v_{i + 1}, a_i)  \in E $ for $ i = 1, \dots, r - 1$.
\end{enumerate}
\end{definition}

\begin{remark}
Notice that if $ u, v \in Q $ are distinct nodes, then $ (u, v) $ trivially precedes $ (u, v) $ itself.
\end{remark}

\begin{lemma}\label{lem:precedesmaximum}
Let $ G = (V, E) $ be a graph, and let $ u, v \in V $ be distinct nodes. Then, there exists a co-lex relation containing $ (u, v) $ if and only if for all  pairs $ (u',v') $ preceding $ (u,v) $ it holds $ \lambda(u')  \; \angle \; \lambda(v') $. In this case, there exists the \emph{minimum} co-lex relation containing $ (u, v) $, that is, a co-lex relation containing $ (u, v) $ refined by every co-lex relation containing $ (u, v) $.
\end{lemma}

\begin{proof}
$ (\Rightarrow) $ Let $ R $ be a co-lex relation containing $ (u, v) $. Assume that $ (u',v') $ precedes $ (u,v) $. We must prove that $ \lambda(u')  \; \angle \; \lambda(v') $. Let $ u_1, \dots, u_r $ and $ v_1, \dots, v_r $ be nodes like in Definition \ref{def:pairprecedes}. From Axiom 2 it follows $ (u_{r - 1}, v_{r - 1}) \in R $, then again by Axiom 2 we obtain $ (u_{r - 2}, v_{r - 2}) \in R $, and so on, until we obtain $ (u', v') \in R $. By Axiom 1 we conclude $ \lambda(u')  \; \angle \; \lambda(v') $.

$ (\Leftarrow) $ Consider a stack that only contains $ (u, v) $ at the beginning. Now, process the element in the stack as follows. Pick $ (u_1, v_1) $ in the stack, remove it from the stack and add to the stack all the pairs $ (u'_1, v'_1) $ of distinct nodes that have not previously been in the stack such that for some $ a \in \Sigma $ it holds $ (u'_1, u_1, a) \in E $ and $ (v'_1, v_1, a) \in E $. Process all the elements in the stack until the stack gets empty (which at some point happens because pairs of states are processed at most once), and let $ R $ be the reflexive closure of the relation obtained by considering all pairs of states that at some point have been in the stack. Let us prove that $ R $ is a co-lex order (and in particular $ (u, v) \in R $). It is immediate to show by induction that all elements that go into the stack precede $ (u, v) $, so by our assumption we have $ \lambda (u) \; \angle \; \lambda (v) $, which proves Axiom 1. Finally, Axiom 2 follows by the rule according to which elements are added to the stack.

Lastly, if there exists a co-lex relation containing $ (u, v) $, then there exists the minimum co-lex relation containing $ (u, v) $, which is simply the relation $ R $ built in $ (\Leftarrow) $: indeed, all elements added to the stack must be in every co-lex relation containing $ (u, v) $ by Axiom 2. \qed
\end{proof}

\begin{corollary}\label{cor:maximumcomputation}
Let $ G = (V, E) $ be a graph, and let $ u, v \in V $ be distinct nodes. Then:
\begin{equation*}
u <_G v  ~ \Leftrightarrow \text{for all  pairs $ (u',v') $ preceding $ (u,v) $ it holds $ \lambda(u')  \; \angle \; \lambda(v') $}.
\end{equation*}
\end{corollary}

\begin{proof}
$ (\Rightarrow) $ Since $ (u, v) $ is contained in a co-lex relation on $ G $ (namely, $ \le_G $), the conclusion follows from Lemma \ref{lem:precedesmaximum}.

$ (\Leftarrow) $ By Lemma \ref{lem:precedesmaximum} $ (u, v) $ is contained in a co-lex order on $ G $, and so also in the maximum co-lex order $ \le_G $.
\end{proof}

\begin{theorem}\label{theor:computemaximumrelation}
Let $ G = (V, E) $ be a graph. Then, $ \le_G $ can be computed in $ O(|E|^2) $ time.
\end{theorem}

\begin{proof}
Consider the graph $ \mathcal{G} = (\mathcal{V}, \mathcal{E}) $, where $ \mathcal{V} = \{(u, v) \in V \times V | u \not = v \} $ and $ \mathcal{E} = \{((u', v'), (u, v)) \in \mathcal{V} | (u', u, a), (v', v, a) \in E \text{ for some $ a \in \Sigma $} \} $. First, mark all $ (u, v) \in \mathcal{V} $ such that $ \lambda (u) \; \angle \; \lambda (v) $ does not hold true (the property "$ \lambda (u) \; \angle \; \lambda (v) $" can be checked in constant time because one only needs to compare the largest element in $ \lambda (u) $ and the smallest element in $ \lambda (v) $). Then, mark all nodes in $ \mathcal{V} $ reachable by a marked node. Notice that at the end a pair $ (u, v) $ is marked if and only if there exists a pair $ (u', v) $ preceding $ (u, v) $ for which $ \lambda (u') \; \angle \; \lambda (v') $ does not hold true, if and only if it holds $ u \not <_G v $ (by Corollary \ref{cor:maximumcomputation}). As a consequence, $ \le_G $ is the reflexive closure of the relation consisting of all non-marked nodes in $ \mathcal{V} $. Notice that $ \le_G $ can be computed in $ O(|E|^2) $ time because $ |\mathcal{E} | \le |E|^2 $ and nodes in $ \mathcal{V} $ can be marked by means of a graph traversal. \qed
\end{proof}

\section{Quotienting a preorder}

As stated in Section \ref{sec:contributions}, we aim to build a quotient graph that captures all information required for pattern matching. Broadly speaking, we will construct the quotient graph starting from a co-lex preorder $ \le $ on $ V $ and considering the \emph{partial} order $ (V /_\le, \le^\sim) $. In this section, we present some preliminary results that will be useful in the following.

\begin{lemma}\label{lem:widthquotientequal}
Let $ (V, \le) $ be a preorder. Then, the width of the partial order $ (V /_\le, \le^\sim) $ is equal to the width of $ (V, \le) $.
\end{lemma}

\begin{proof}
Let $ m_1 $ be the width of $ (V /_\le, \le^\sim) $ and let $ m_2 $ be the width of $ (V, \le) $. We must prove that $ m_1 = m_2 $. On the one hand, if $ \{U_i \}_{i = 1}^{m_1} $ is a $ \le^\sim$-chain decomposition of $ V /_\le $, then $ \{V_i \}_{i = 1}^{m_1} $ is a $ \le $-chain decomposition of $ V $, where $ V_i $ is the union of all elements of $ V $ being in some $ \sim_\le $-class of $ U_i $. This proves that $ m_2 \le m_1 $. On the other hand, if $ \{V_i \}_{i = 1}^{m_2} $ is a $ \le $-chain decomposition of $ V $, then $ \{U_i \}_{i = 1}^{m_2} $ is a cover of $ V /_\le $, where $ U_i = \{[v]_\le | v \in V_i \} $, and each $ U_i $ is a $ \le^\sim $ chain, so by extracting an arbitrary partition from the cover we obtain a $ \le^\sim$-chain decomposition of $ V /_\le $ of cardinality at most $ m_2 $. This proves that $ m_1 \le m_2 $. \qed
\end{proof}

Incidentally, Lemma \ref{lem:widthquotientequal} is the start point for proving a Dilworth theorem-like for preorders. \emph{Dilworth theorem} \cite{dilworth2009decomposition} states that the width of a partial order is equal to the maximum size of an antichain. The same results holds true for preorders. This results is likely to have been implicitly proved previously, but since we did not find a statement for for preorders in the literature, we provide an explicit proof.

\begin{theorem}[Dilworth theorem for preorders]\label{theor:Dilworthpreorders}
Let $ (V, \le) $ be a preorder. Then, the width of $ (V, \le) $ is equal to the maximum size of a $ \le $-antichain. 
\end{theorem}

\begin{proof}
Consider the partial order $ (V /_\le, \le^\sim) $. By Dilworth theorem for partial orders \cite{dilworth2009decomposition}, the width of $ (V /_\le, \le^\sim) $ is equal to the the maximum size of a $ \le^\sim $-antichain in $ V /_\le $. The theorem will follow if we prove that the width of $ (V /_\le, \le^\sim) $ is equal to the width of $ (V, \le) $, and the maximum size of a $ \le^\sim $-antichain in $ V /_\le $ is equal to the maximum size of a $ \le $-antichain in $ V $. The first statement is Lemma \ref{lem:widthquotientequal}, so we only have to prove the second statement.

Let $ M_1 $ be the maximum size of an $ \le^\sim $-antichain in $ V /_\le $ and let $ M_2 $ be the maximum size of a $ \le $-antichain in $ V $. We must prove prove that $ M_1 = M_2 $. On the one hand, if $ \{[v_1]_\le, \dots, [v_{M_1}]_\le \} $ is a $ \le^\sim $-antichain in $ V /_\le $, then $ \{v_1, \dots, v_{M_1} \} $ is a $ \le $-antichain in $ V $. This prove that $ M_1 \le M_2 $. On the other hand, if $ \{v_1, \dots, v_{M_2} \} $ is a $ \le $-antichain in $ V $, then the elements of the antichain are in pairwise distinct $ \sim_\le $-classes, so $ \{[v_1]_\le, \dots, [v_{M_2}]_\le \} $ is a $ \le^\sim $-antichain in $ V /_\le $ and, in fact, it has cardinality $ M_2 $. This proves that $ M_2 \le M_1 $. \qed
\end{proof}

Let us prove a simple result relating convexity and quotients: every convex set is the union of some $ \sim_\le $-classes. This result is crucial for showing that without loss of generality we can perform pattern matching on the quotient graph.

\begin{lemma}\label{lem:classesareinconvex}
Let $ (V, \le) $ be a preorder, and let $ U \subseteq V $ be $ \le $-convex. If $ v \in U $, then $ [v]_\sim \subseteq U $. In other words, every $ \le $-convex set is the union of some $ \sim_\le $-classes.
\end{lemma}

\begin{proof}
Assume that $ u \in [v]_\sim  $. We must prove that $ u \in U $. We know that $ v \in U $, $ v \le u $ and $ u \le v $, so we conclude $ u \in U $ because $ U $ is $ \le $-convex. \qed
\end{proof}

More generally, we can prove that there is a natural 1-1 correspondence between $ \le $-convex sets in $ V $ and $ \le^\sim $-convex sets in $ V /_\le $.

\begin{lemma}[Correspondence theorem - convex sets]\label{lem:quotientcorrespondenceconvex}
Let $ (V, \le) $ be a preorder. Let $ \mathcal{U} $ be the family of all $ \le $-convex sets in $ V $, and let $ \mathcal{U}_\le $ be the family of all $ \le^\sim $-convex sets in $ V /_\le $. Define:
\begin{align*}
  \phi: \mathcal{U} &\to \mathcal{U}_\le \\
  U &\mapsto \{[v]_\le | v \in U \}.
\end{align*}
Then, $ \phi $ is a bijective function, with inverse:
\begin{align*}
  \psi: \mathcal{U}_\le &\to \mathcal{U} \\
  U_\le &\mapsto \{v \in V | [v]_\le \in U_\le \}.
\end{align*}
\end{lemma}

\begin{proof}
First, let us prove that that $ \phi $ and $ \psi $ well-defined.
\begin{enumerate}
    \item Let us prove that if $ U $ is $ \le $-convex, then $ U_\le = \{[v]_\le | v \in U \} $ is $ \le^\sim $-convex. Assume that $ [u]_\le, [v]_\le, [z]_\le \in V /_\le $ satisfy $ [u]_\le, [z]_\le \in U_\le $, $ [u]_\le \le^\sim [v]_\le $ and $ [v]_\le \le^\sim [z]_\le $. We must prove that $ [v]_\le \in U_\le $. From $ [u]_\le \le^\sim [v]_\le $ and $ [v]_\le \le^\sim [z]_\le $ it follows $ u \le v $ and $ v \le z $. Moreover, from $ [u]_\le, [z]_\le \in U /_\le $ and Lemma \ref{lem:classesareinconvex} it follows $ u, z \in U $. Since $ U $ is $ \le $-convex, we conclude $ v \in U $, and so $ [v]_\le \in U_\le $.
    \item Let us prove that if $ U_\le $ is $ \le^\sim $-convex, then $ U = \{v \in V | [v]_\le \in U_\sim \} $ is $ \le $-convex. Assume that $ u, v, z \in V $ satisfy $ u, z \in U $, $ u \le v $ and $ v \le z $. We must prove that $ v \in U $. From $ u \le v $ and $ v \le z $ it follows $ [u]_\le \le^\sim [v]_\le $ and $ [v]_\le \le^\sim [z]_\le $. Moreover, from $ u, z \in U $ it follows $ [u]_\le, [z]_\le \in U_\le $. Since $ U_\le $ is $ \le^\sim $-convex, we conclude $ [v]_\le \in U_\le $, and so $ v \in U $.
\end{enumerate}
Now, we are only left with proving that $ \psi \circ \phi = id_\mathcal{U} $ and $ \phi \circ \psi = id_\mathcal{U_\le} $. We have:
\begin{equation*}
    (\psi \circ \phi)(U) = \psi(\{[v]_\le | v \in U \}) = \{v' \in U | \text{  $ [v']_\le = [v]_\le $ for some $ v \in U $} \} =  U
\end{equation*}
where ($ \subseteq $) in the last equality follows from Lemma \ref{lem:classesareinconvex}. Finally:
\begin{equation*}
    (\phi \circ \psi)(U_\le) = \phi (\{v \in V | [v]_\le \in U_\le \}) = \{[v']_\le | [v']_\le \in U_\le \} = U_\le.
    \end{equation*}
\qed
\end{proof}

\section{The quotient graph}

We can now define our quotient graph.

\begin{definition}\label{def:quotientcolex}
Let $ G = (V, E) $ be a graph, and let $ \le $ be a co-lex preorder on $ G $. Define $ G /_\le  = (V /_\le, E /_\le ) $ by:
\begin{enumerate}
    \item $ V /_\le = \{[v]_\le | v \in V \} $;
    \item $ E /_\le = \{([u]_\le, [v]_\le, a) | (u', v', a) \in E \text{ for some $ u' \in [u]_\le $ and $ v' \in [v]_\le $}   \}  $.
\end{enumerate}
\end{definition}

\begin{remark}\label{rem:propertiescolexquotient}
(1) If $ u, v \in V $ are distinct nodes such that $ [u]_\le = [v]_\le $, then $ \lambda (u) = \lambda (v) $ and $ |\lambda (u)| = |\lambda (v)| = 1 $ by Remark \ref{rem:initialproperty}. (2) If $ [u]_\le = [v]_\le $, then $ \lambda (u) = \lambda (v) $. Indeed, if $ u $ and $ v $ are distinct nodes, the conclusion follows from the first point, otherwise the conclusion is trivial (in this case, if $ [v]_\le = \{v \} $, then $ \lambda (v) $ may have cardinality larger than one. (3) For every $ v \in V $, it holds $ \lambda (v) = \lambda ([v]_\le) $, where $ \lambda (v) $ refers to $ G $ and $ \lambda ([v]_\le) $ refers to $ G /_\le $. Indeed, if $ a \in \Sigma \cap \lambda (v) $, then there exists $ u \in V $ such that $ (u, v, a) \in E $, so $ ([u]_\le, [v]_\le, a) \in E /_\sim $ and $ a \in \lambda (v]_\sim) $; conversely, if $ a \in \Sigma \cap \lambda (v]_\le) $, then there exist $ u', v' \in V $ such that $ (u', v', a) \in E $ and $ [v']_\le = [v]_\sim $, so $ a \in \lambda (v') $ and, by the second point, $ a \in \lambda (v) $.
\end{remark}

Let us prove that $ G /_\le $ enjoys a number of properties. (1) If a node of $ G /_\le $ has been obtained by collapsing two or more nodes of $ G $, then that node has at most one ingoing edge in $ G /_\le $ (which is possibly a self-loop). (2) $ \le^\sim $ is a co-lex order on $ G /_\le $. (3) The graph $ G /_{\le_G }$ always admits the maximum co-lex order (recall that in general a graph does not admit the maximum co-lex order). More precisely, the maximum co-lex order is $ \le_G^\sim $ (the partial order on $ V /_{\le_G} $ induced by $ \le_G $), which is also the maximum co-lex relation on $ G /_{\le_G} $. Notice that $ G /_{\le_G }$ is well-defined because $ \le_G $ is a co-lex preorder by Lemma \ref{lem:existencemaximumcolexrelation}.

We prove the first property in Lemma \ref{lem:quotientjustoneedgeentering}. We need a preliminary result.

\begin{lemma}\label{lem:edgesbetweenclasses}
Let $ G = (V, E) $ be a graph, and let $ \le $ be a co-lex preorder on $ G $. Assume that $ u, v \in V $ are (non necessarily distinct) nodes such that $ [u]_\le = [v]_\le $ and $ |[u]_\le| = |[v]_\le| \ge 2 $. If $ (u', u, a), (v', v, a) \in E $, then $ [u']_\le = [v']_\le $.
\end{lemma}

\begin{proof}
We distinguish two cases.
\begin{enumerate}
    \item Assume that $ u \not = v $. From $ [u]_\le = [v]_\le $ we obtain $ u < v $ and $ v < u $, hence Axiom 2 applied to $ (u', u, a) $ and $ (v', v, a) $ implies $ u' \le v' $ and $ v' \le u' $, so $ [u']_\le = [v']_\le $.
    \item Assume that $ u = v $. Since $ |[u]_\le| \ge 2 $, then there exists $ z \in V $ such that  $ u \not = z $ and $ [u]_\le = [z]_\le $. From Remark \ref{rem:propertiescolexquotient}, we know that $ \lambda (u) = \lambda (v) = \lambda (z) = \{a \} $, so there exists $ z' \in V $ such that $ (z', z, a) \in E $. From $ [u]_\le = [z]_\le $ we obtain $ u < z $ and $ z < u $, hence Axiom 2 applied to $ (u', u, a) $ and $ (z', z, a) $ implies $ u' \le z' $ and $ z' \le u' $, and Axiom 2 applied to $ (v', v, a) $ and $ (z', z, a) $ implies $ v' \le z' $ and $ z' \le v' $. Hence, $ [u']_\le = [z']_\le $ and $ [v']_\le = [z']_\le $, and so $ [u']_\le = [v']_\le $.
\end{enumerate}
\qed
\end{proof}

\begin{lemma}\label{lem:quotientjustoneedgeentering}
Let $ G = (V, E) $ be a graph, and let $ \le $ be a co-lex preorder on $ G $. If $ [v]_\le \in V /_\le $ is such that $ |[v]_\le| \ge 2 $, then there exists at most one edge entering $ [v]_\le $ in $ G /_\le $.
\end{lemma}

\begin{proof}
Since $ |[v]_\le| \ge 2 $, by Remark \ref{rem:propertiescolexquotient} we have $ |\lambda ([v]_\le)| = 1 $. If $ \lambda ([v]_\le) = \{\# \} $, then there is no edge entering $ [v]_\le $ in $ G /_\le $. Now, assume $ \lambda ([v]_\le) = \{a \} $, with $ a \in \Sigma $, and let $ ([v']_\le, [v]_\le, a), ([v'_1]_\le, [v]_\le, a) \in E /_\le $. We must prove that these two edges are actually the same edge, that is, $ [v']_\le = [v'_1]_\le $. Since $ ([v']_\le, [v]_\le, a), ([v'_1]_\le, [v]_\le, a) \in E /_\le $, then there exist $ u, u_1, u', u'_1 \in V  $ such that $ [v]_\le = [u]_\le = [u_1]_\le $, $ [v']_\le = [u']_\le $, $ [v'_1]_\le = [u'_1]_\le $, $ (u', u, a) \in E $ and $ (u'_1, u_1, a) \in E $. Since $ [u]_\le = [u_1]_\le $ and $ |[u]_\le| = |[u_1]_\le| = |[v]_\le| \ge 2 $, then from Lemma \ref{lem:edgesbetweenclasses} we obtain $ [u']_\le = [u'_1]_\le $, so from $ [v']_\le = [u']_\le $ and $ [v'_1]_\le = [u'_1]_\le $ we conclude $ [v']_\le = [v'_1]_\le $. \qed
\end{proof}

Next, we prove that $ \le^\sim $ is a co-lex order on $ G /_\le $.

\begin{lemma}\label{lem:quotientcolex}
Let $ G = (V, E) $ be a graph, and let $ \le $ be a co-lex preorder on $ G $. Then, $ \le^\sim $ is a co-lex order on $ G /_\le $, and the width of $ \le^\sim $ is equal to the width of $ \le $.
\end{lemma}

\begin{proof}

Let us prove that $ \le^\sim $ is a co-lex order on $ G /_\le $. We know that $ \le^\sim $ is a partial order, so we only have to prove that it satisfies Axiom 1 and Axiom 2.

Let us prove Axiom 1. Assume that $ [u]_\le, [v]_\le \in V /_\le $ satisfy $ [u]_\le <^\sim [v]_\sim $. We must prove that $ \lambda ([u]_\le) \; \angle \; \lambda (v]_\le) $. By the definition of $ \le^\sim $ we have $ u < v $, so by Axiom 1 applied to $ \le $ we conclude $ \lambda (u) \; \angle \; \lambda (v) $. The conclusion follows, because by Remark \ref{rem:propertiescolexquotient} we have $ \lambda ([u]_\le) = \lambda (u) $ and $ \lambda ([v]_\le) = \lambda (v) $.

Let us prove Axiom 2. Assume that $ ([u']_\le, [u]_\le, a), ([v']_\le, [v]_\le, a) \in E /_\le $ satisfy $ [u]_\le <^\sim [v]_\le $. We must prove that $ [u']_\le \le^\sim [v']_\le $. Since $ ([u']_\le, [u]_\le, a) \in E /_\le $, then there exist $ u'_1, u_1 \in V $ such that $ (u'_1, u_1, a) \in E $, $ [u'_1]_\le = [u']_\le $ and $ [u_1]_\le = [u]_\le $. Analogously, $ ([v']_\le, [v]_\le, a) \in E /_\le $ implies that there exist $ v'_1, v_1 \in V $ such that $ (v'_1, v_1, a) \in E $, $ [v'_1]_\le = [v']_\le $ and $ [v_1]_\le = [v]_\le $. From $ [u]_\le <^\sim [v]_\le $ we obtain $ u_1 < v_1 $, hence by Axiom 2 applied to $ \le $ we conclude $ u'_1 \le v'_1 $, which implies $ [u']_\le \le^\sim [v']_\le $.

Lastly, $ \le^\sim $ and $ \le $ have the same width by Lemma \ref{lem:widthquotientequal}. \qed
\end{proof}

Let us prove that $ \le_G^\sim $ is the maximum co-lex relation and the maximum co-lex order on $ G /_{\le_G} $.

\begin{lemma}[Correspondence theorem - co-lex relations]\label{lem:correspondencequotientcolexrelations}
Let $ G= (V, E) $ be a graph, and let $ \le $ be a co-lex preorder on $ G $. Let $ \mathcal{C}_\le $ the set of all co-lex relations on $ G /_\le $, and let $ \mathcal{C} $ be set of all co-lex relations $ R $ on $ G $ such that, if $ (u, v) \in R $, $ [u]_\le = [u']_\le $ and $ [v]_\le = [v']_\le $, then $ (u', v') \in R $. Define:
\begin{align*}
  \rho: \mathcal{C}_\le &\to \mathcal{C} \\
  R^\le &\mapsto \{(u, v) \in V \times V | ([u]_\le, [v]_\le) \in R^\le \}
\end{align*}
Then, $ \rho $ is a bijective function, with inverse:
\begin{align*}
  \sigma: \mathcal{C} &\to \mathcal{C}_\le \\
  R &\mapsto \{([u]_{\le}, [v]_\le) \in V /_\le \times V /_\le | (u, v) \in R \}.
\end{align*}
In particular, $ \sigma (\le) $ is equal to $ \le^\sim $.
\end{lemma}

\begin{proof}
First, let us prove that $ \rho $ and $ \sigma $ are well-defined.
\begin{enumerate}
    \item Let us prove that if $ R^\le $ is a co-lex relation on $ G /_\le $, then $ R = \{(u, v) \in V \times V | (u, v) \in R \} $ is a co-lex relation on $ G $ which belongs to $ \mathcal{C} $. Clearly, $ R $ is reflexive because $ R^\le $ is reflexive. Let us prove Axiom 1. Assume that $ (u, v) \in R  $, with $ u \not = v $. We must prove that $ \lambda (u) \; \angle \; \lambda (v) $. It must be $ ([u]_\le, [v]_\le) \in R  $. If $ [u]_\le \not = [v]_\le  $, then from Axiom 1 applied to $ R^\le $ we obtain $ \lambda ([u]_\le) \; \angle \; \lambda ([v]_\le ) $, and we conclude $ \lambda (u) \; \angle \; \lambda (v) $ because $ \lambda (u) = \lambda ([u]_\le) $ and $ \lambda (v) = \lambda ([v]_\le) $ by Remark \ref{rem:propertiescolexquotient}. If $ [u]_\le = [v]_\le $, then again from Remark \ref{rem:propertiescolexquotient} we obtain $ \lambda (u) = \lambda (v) $ and $ |\lambda (u)| = |\lambda (v)| = 1 $, so again $ \lambda (u) \; \angle \; \lambda (v) $ (see Remark \ref{rem:propertiesangletry}). Let us prove Axiom 2. Assume that $ (u', u, a), (v', v, a) \in E $, with $ (u, v) \in R $ and $ u \not = v $. We must prove that $ (u', v') \in R $. We have $ ([u']_\le, [u]_\le, a), ([v']_\le, [v]_\le, a) \in E /_\le  $ and $ ([u]_\le, [v]_\le) \in R^\le $. If $ [u]_\le \not = [v]_\le  $, then from Axiom 2 applied to $ R^\le $ we obtain $ ([u']_\le, [v']_\le) \in R^\le $, and so $ (u', v') \in R $. If $ [u]_\le = [v]_\le  $, then we have $ |[u]_\le| = |[v]_\le| \ge 2 $ (because $ u \not = v $), so by Lemma \ref{lem:quotientjustoneedgeentering} we have $ [u']_\le = [v']_\le $, hence trivially $ ([u']_\le, [v']_\le) \in R^\le $ and in particular $ (u', v') \in R $. Lastly, $ R $ belongs to $ \mathcal{C} $ because if $ (u, v) \in R $, $ [u]_\le = [u']_\le $ and $ [v]_\le = [v']_\le $, then $ ([u']_\le, [v']_\le) = ([u]_\le, [v]_\le) \in R^\le $ and so $ (u', v') \in R $.
    \item Let us prove that if $ R $ a co-lex relation in $ \mathcal{C} $, then $ R^\le = \{([u]_{\le}, [v]_\le) \in V /_\le \times V /_\le | (u, v) \in R \} $ is a well-defined co-lex relation on $ G /_\le $. First, $ R^\le $ is well-defined (that is, if $ [u']_\le = [u]_\le $, $ [v']_\le = [v]_\le $ and $ (u, v) \in R $, then $ (u', v') \in R $) because $ R $ belongs to $ \mathcal{C} $. Clearly, $ R^\le $ is reflexive because $ R $ is reflexive. Let us prove Axiom 1. Assume that $ ([u]_\le, [v]_\le) \in R^\le $, with $ [u]_\le \not = [v]_\le $. We must prove that $ \lambda ([u]_\le) \; \angle \; \lambda ([v]_\le) $. It must be $ (u, v) \in R $, with $ u \not = v $, so from Axiom 1 applied to $ R $ we obtain $ \lambda (u) \; \angle \; \lambda (u) $, and we conclude $ \lambda ([u]_\le) \; \angle \; \lambda ([v]_\le) $ because $ \lambda (u) = \lambda ([u]_\le) $ and $ \lambda (v) = \lambda ([v]_\le) $ by Remark \ref{rem:propertiescolexquotient}. Let us prove Axiom 2. Assume that $ ([u']_\le, [u]_\le, a), ([v']_\le, [v]_\le, a) \in E /_\le $, with $ ([u]_\le, [v]_\le) \in R^\le $ and $ [u]_\le \not = [v]_\le $. We must prove that $ ([u']_\le, [v']_\le) \in R^\le $. There must exist $ u_1, v_1, u'_1, v'_1 \in V $ such that $ [u_1]_\le = [u]_\le $, $ [v_1]_\le = [v]_\le $, $ [u'_1]_\le = [u']_\le $, $ [v'_1]_\le = [v']_\le $, $ (u'_1, u_1, a) \in E $ and $ (v'_1, v_1, a) \in E $. From $ ([u]_\le, [v]_\le) \in R^\le $, $ [u_1]_\le = [u]_\le $ and $ [v_1]_\le = [v]_\le $ it follows $ (u_1, v_1) \in R $, where $ u_1 \not = v_1 $ (because  $ [u]_\le \not = [v]_\le $). From Axiom 2 applied to $ R $ we obtain $ (u'_1, v'_1) \in R $, hence from $ [u'_1]_\le = [u']_\le $ and $ [v'_1]_\le = [v']_\le $ we conclude $ ([u']_\le, [v']_\le) \in R^\le $.
\end{enumerate}

Next, let us prove that  $ \sigma \circ \rho = id_{\mathcal{C}_\le} $ and $ \rho \circ \sigma = id_{\mathcal{C}} $. We have:
\begin{equation*}
\begin{split}
    (\sigma \circ \rho)(R^\le) & = \sigma (\{(u, v) \in V \times V | ([u]_\le, [v]_\le) \in R^\le \}) = \\
    & = \{([u]_{\le}, [v]_\le) \in V /_\le \times V /_\le | ([u]_\le, [v]_\le) \in R^\le \} = R^\le
\end{split}
\end{equation*}
and:
\begin{equation*}
\begin{split}
    (\rho \circ \sigma)(R) & = \rho (\{([u]_{\le}, [v]_\le) \in V /_\le \times V /_\le | (u, v) \in R \}) = \\
    & = \{(u, v) \in V \times V | (u, v) \in R \} = R.
\end{split}
\end{equation*}

Lastly, notice that $ \le $ belongs to $ \mathcal{C} $, because if $ u \le v $, $ [u]_\le = [u']_\le $ and $ [v]_\le = [v']_\le $, then $ u' \le u \le v \le v' $. By the definition of $ \sigma $, we conclude that $ \sigma (\le) $ is equal to $ \le^\sim $. \qed
\end{proof}

\begin{corollary}\label{cor:quotientgraphmaximumcolexorder}
Let $ G = (V, E) $ be a graph. Then, $ \le^\sim_G $ is the maximum co-lex relation and the maximum co-lex order on $ G /_{\le_G} $.
\end{corollary}

\begin{proof}
By Lemma \ref{lem:quotientcolex} we know that $ \le^\sim_G $ is a co-lex order on $ G /_{\le_G} $, so we only have to prove that $ \le^\sim_G $ is the maximum co-lex relation on $ G /_{\le_G} $. Let $ R^{\le_G} $ be a co-lex relation on $ G /_{\le_G} $ and assume that $ ([u]_{\le_G}, [v]_{\le_G}) \in R^{\le_G} $. We must prove that $ [u]_{\le_G} \le^\sim_G [v]_{\le_G} $. By Lemma \ref{lem:correspondencequotientcolexrelations} we know that $ R = \rho (R^{\le_G}) $ is a co-lex relation on $ G $, and $ (u, v) \in R $. Since $ \le_G $ is the maximum co-lex relation on $ G $, we obtain $ u \le_G v $, so we conclude $ [u]_{\le_G} \le^\sim_G [v]_{\le_G} $. \qed
\end{proof}

\section{Indexing for pattern matching}

Recall that in \cite{cotumaccio2021indexing} it was showed how to index a graph by means of a co-lex order. However, determining a co-lex order of minimum width is a hard problem \cite{cotumaccio2021indexing}. On the other hand, Corollary \ref{cor:quotientgraphmaximumcolexorder} ensures that $ G /_{\le_G} $ always admits the maximum co-lex order (which has minimum width), and it can be determined in polynomial time by Theorem \ref{theor:computemaximumrelation}. As a consequence, we have overcome the hardness of determining a co-lex order of minimum width of an \emph{arbitrary} graph if we show that we can answer pattern matching queries on $ G $ \emph{by answering a query on $ G /_{\le_G} $}. This in indeed the purpose of the following lemma. Intuitively, if we start from a $ \le_G $-convex set $ U $ of nodes in $ G $, we can obtain the $ \le $-convex set of nodes that can be reached through a string $ \alpha $ by (1) passing to the quotient, (2) obtaining the $ \le_G^\sim $-convex set of nodes that can reached through $ \alpha $ in $ G /_{\le_G} $, and (3) going back to $ G $.

\begin{lemma}[Correspondence theorem - path coherence]\label{lem:intervalsgointointervals}
Let $ G = (V, E) $ be a graph, and let $ \le $ be a co-lex preorder on $ G $. Let $ \alpha \in \Sigma^* $. Let $ \mathcal{U} $ be the family of all $ \le $-convex sets in $ V $, and let $ \mathcal{U}_\le $ be the family of all $ \le^\sim $-convex sets in $ V /_\le $. Let:
\begin{align*}
  \theta_\alpha: \mathcal{U} &\to \mathcal{U} \\
\end{align*}
be the function such that if $ U \in \mathcal{U} $, then $ \theta_\alpha (U) $ is the set of all nodes of $ G $ that can be reached from $ U $ by following edges whose labels, when concatenated, yield $ \alpha $. Moreover, let:
\begin{align*}
  \theta^\le_\alpha: \mathcal{U}_\le &\to \mathcal{U}_\le \\
\end{align*}
be the function such that if $ U_\le \in \mathcal{U}_\le $, then $ \theta^\le_\alpha (U_\le) $ is the set of all nodes of $ G /_\le $ that can be reached from $ U_\le $ by following edges whose labels, when concatenated, yield $ \alpha $. Let $ \phi $ and $ \psi $ the functions defined in Lemma \ref{lem:quotientcorrespondenceconvex}. Then (see Figure \ref{fig:commutativepathcoherence}):
\begin{align*}
    \theta_\alpha \circ \psi  &= \psi \circ \theta^\le_\alpha \\
    \phi \circ \theta_\alpha &= \theta^\le_\alpha \circ \phi.
\end{align*}

\begin{comment}
Let $ G = (V, E) $ be a graph, and let $ \le $ be a co-lex preorder on $ G $. Let $ \alpha \in \Sigma^* $, and let $ U_\sim \subseteq V_\sim $ be $ \le^\sim $-convex. Let $ U'_\sim \subseteq V_\sim $ the set of all nodes all nodes of $ G_\sim $ that can be reached from $ U_\sim $ by following edges whose labels, when concatenated, yield $ \alpha $ (the set $ U'_\sim $ is $ \le^\sim $-convex by Lemma \ref{lem:quotientcolex} and Lemma \ref{lem:pathcoherencecolexrelations}). Define:
\begin{align*} 
U &= \{v \in V | [v]_\sim \in U_\sim    \} \\
U' &= \{v \in V | [v]_\sim \in U'_\sim    \}. \\
\end{align*}
Then, $ U $ and $ U' $ are $ \le $-convex. Moreover, $ U' $ is the set of all nodes all nodes of $ G $ that can be reached from $ U $ by following edges whose labels, when concatenated, yield $ \alpha $.
\end{comment}

\begin{figure}
\centering
\begin{tikzpicture}[baseline= (a).base]
\node[scale=2] (a) at (0,0){
\begin{tikzcd}
\mathcal{U} \arrow[r, "\phi"] \arrow[d, "\theta_\alpha"]
& \mathcal{U}_\le \arrow[d, "\theta^\le_\alpha"] \arrow[l, "\psi", shift left] \\
\mathcal{U} \arrow[r, "\phi"]
& \mathcal{U}_\le \arrow[l, "\psi", shift left]
\end{tikzcd}
};
\end{tikzpicture}
\caption{Lemma \ref{lem:intervalsgointointervals}}\label{fig:commutativepathcoherence}

\end{figure}
\end{lemma}

\begin{proof}
First, the codomains of $ \theta_\alpha $ and $ \theta^\le_\alpha $ are correct by Lemma \ref{lem:pathcoherencecolexrelations}.

Let us prove the first equation. We proceed by induction on $ |\alpha| $. If $ |\alpha| = 0 $, then $ \alpha = \epsilon $, so $ \theta_\alpha = id_{\mathcal{U}} $, $ \theta^\le_\alpha = id_{\mathcal{U}_\le} $ and we conclude $ \theta_\alpha \circ \psi = \psi = \psi \circ \theta^\le_\alpha $. Now, assume $ |\alpha| \ge 1 $. We can write $ \alpha = \alpha' a $, with $ \alpha' \in \Sigma^* $ and $ a \in \Sigma $. By the inductive hypothesis, $ \theta_{\alpha'} \circ \psi  = \psi \circ \theta^\le_{\alpha'} $. Moreover, notice that $ \theta_\alpha = \theta_a \circ \theta_{\alpha'} $ and $ \theta^\le_\alpha = \theta^\le_a \circ \theta^\le_{\alpha'} $. Hence:
\begin{align*}
    \theta_\alpha \circ \psi = \theta_a \circ \theta_{\alpha'} \circ \psi & = \theta_a \circ \psi \circ \theta^\le_{\alpha'}  \\
    \psi \circ \theta^\le_\alpha & = \psi \circ \theta^\le_a \circ \theta^\le_{\alpha'}
\end{align*}
so the conclusion follows if we prove that:
\begin{equation*}
    \theta_a \circ \psi = \psi \circ \theta^\le_a.
\end{equation*}
Fix $ U_\le \in \mathcal{U}_\le $. We have to prove that $ (\theta_a \circ \psi)(U_\le) = (\psi \circ \theta^\le_a)(U_\le) $. We will use that $ \psi $ is the inverse of $ \phi $ (Lemma \ref{lem:quotientcorrespondenceconvex}).

($ \subseteq $) If $ v \in (\theta_a \circ \psi)(U_\le) $, then there exists $ u \in \psi (U_\le) $ such that $ (u, v, a) \in E $. In particular, $ [u]_\le \in U_\le $ and $ ([u]_\le, [v]_\le, a) \in E /_\le $, so $ [v]_\le \in \theta^\le_a (U_\le) $ and we conclude $ v \in (\psi \circ \theta^\le_a)(U_\le) $.

($ \supseteq $) If $ v \in (\psi \circ \theta^\le_a)(U_\le) $, then $ [v]_\le \in \theta^\le_a(U_\le) $, so there exists $ [u]_\le \in U_\le $ such that $ ([u]_\le, [v]_\le, a) \in E /_\le $. In particular, there exist $ u', v' \in V $ such that $ (u', v', a) \in E $, $ [u']_\le = [u]_\le \in U_\le $ and $ [v']_\le = [v]_\le $. This means that $ v' \in (\theta_a \circ \psi)(U_\le) $, and by Lemma \ref{lem:classesareinconvex} we conclude $ v \in (\theta_a \circ \psi)(U_\le) $.

Let us prove the second equation. Since $ \psi $ is the inverse of $ \phi $ (Lemma \ref{lem:quotientcorrespondenceconvex}), then from the first equation we obtain:
\begin{equation*}
    \phi \circ \theta_\alpha = \phi \circ \theta_\alpha \circ \psi \circ \phi = \phi \circ \psi \circ \theta^\le_\alpha \circ \phi = \theta^\le_\alpha \circ \phi.
\end{equation*}
\qed
\end{proof}

\begin{corollary}\label{cor:patternmatchingonquotientisequivalent}
Let $ G = (V, E) $ be a graph, and let $ \le $ be a co-lex preorder on $ G $. Let $ \alpha \in \Sigma^* $. Then the pattern matching problem returns "yes" on input $ G $ and $ \alpha $ if and only if it returns "yes" on input $ G /_{\le} $ and $ \alpha $.
\end{corollary}

\begin{proof}
Using the notation of Lemma \ref {lem:intervalsgointointervals}, the pattern matching problem returns "yes" on input $ G $ and $ \alpha $ if and only if $ \theta_\alpha (V) \not = \emptyset $ (where $ V $ is trivially $ \le $-convex), and it returns "yes" on input $ G /_{\le} $ and $ \alpha $ if and only if $ \theta^\le_\alpha (V /_{\le}) \not = \emptyset $. On the other, by Lemma \ref {lem:intervalsgointointervals} we have $ \theta^\le_\alpha (V /_{\le}) = (\theta^\le_\alpha (\phi (V)) = \phi (\theta_\alpha (V)) $. The conclusion follows, because for every $ U \in \mathcal{U} $ it holds $ U = \emptyset $ if and only if $ \phi (U) = \emptyset $. \qed
\end{proof}

We now recall the main result from \cite{cotumaccio2021indexing} and we adapt it to edge-labeled graphs.

\begin{theorem}\label{theor:partialordersindexinggeneral}
Let $ G = (V, E) $ be a graph, and assume that we are given a co-lex order $ \le $ on $ G $ of minimum width $ p $. Then, there exists a data structure of $|E|(\lceil \log|\Sigma|\rceil + \lceil\log p\rceil + 2)\cdot (1+o(1)) + |V|\cdot (1+o(1))$ bits, which can be built starting from $ G $ and $ \le $ in $ O(|V|^{5 / 2 }) $ time, such that, given a pattern $ P \in \Sigma^* $ and a $ \le $-convex set $ U $, in $O(|P| \cdot p^2 \cdot \log(p\cdot |\Sigma|)) $ time returns the $ \le $-convex set of all nodes in $ V $ that can be reached from $ U $ by following edges whose labels, when concatenated, yield $ P $.
\end{theorem}

\begin{proof}
This is essentially \cite[Thm 4.2]{cotumaccio2021indexing}. The only difference is that the cited theorem refers to node-labeled graphs. However, since we have proved that path coherence also holds for edge-labeled graphs (Lemma \ref{lem:pathcoherencecolexrelations}), one readily checks that the same proof also works in our more general setting. Notice also that \cite[Thm 4.2]{cotumaccio2021indexing} refers to automata, so in our graph setting we need a slightly smaller amount of bits because we do not need to store final states. \qed
\end{proof}

\begin{theorem}\label{theor:partialorderspatternmatching}
Let $ G = (V, E) $ be a graph, and assume that we are given a co-lex order $ \le $ on $ G $ of minimum width $ p $. Then, there exists a data structure of $|E|(\lceil \log|\Sigma|\rceil + \lceil\log p\rceil + 2)\cdot (1+o(1)) + |V|\cdot (1+o(1))$ bits, which can be built starting from $ G $ and $ \le $ in $ O(|V|^{5 / 2 }) $ time, that solves the pattern-matching problem in $O(|P| \cdot p^2 \cdot \log(p\cdot |\Sigma|)) $ time, where $ P \in \Sigma^* $ is the pattern.
\end{theorem}

\begin{proof}
The conclusion follows from Theorem \ref{theor:partialordersindexinggeneral} by letting $ U = V $, which is trivially $ \le $-convex. \qed
\end{proof}

Theorems \ref{theor:partialordersindexinggeneral} and \ref{theor:partialorderspatternmatching} simply generalize the results in \cite{cotumaccio2021indexing} from node-labeled graphs to edge-labeled graphs. However, building the data structure in Theorems \ref{theor:partialordersindexinggeneral} and \ref{theor:partialorderspatternmatching} does not require only $ O(|V|^{5 / 2 }) $ time, because the theorems assume that we are given a co-lex order $ \le $ on $ G $ of minimum width $ p $, and determining such a co-lex order is a hard problem \cite{cotumaccio2021indexing}.

We can now overcome this limitation by passing to the quotient graph. Here is our main result.

\begin{theorem}\label{theor:patternmatchingpreorders}
Let $ G = (V, E) $ be a graph, and let $ q $ be the the width of $ \le_G $. Then, there exists a data structure of $|E /_{\le_G}|(\lceil \log|\Sigma|\rceil + \lceil\log q\rceil + 2)\cdot (1+o(1)) + |V /_{\le_G}|\cdot (1+o(1))$ bits, which can be built starting from $ G $  in $ O(|E|^2 + |V /_{\le_G}|^{5 / 2 }) $ time, that solves the pattern-matching problem in $O(|P| \cdot q^2 \cdot \log(q\cdot |\Sigma|)) $ time, where $ P \in \Sigma^* $ is the pattern.
\end{theorem}

\begin{proof}
Compute $ \le_G $ in $ O(|E|^2) $ time (Theorem \ref{theor:computemaximumrelation}). Build the graph $ G /_{\le_G} $ by a graph traversal. By Lemma \ref{lem:quotientcolex} we know that $ \le^\sim_G $ is a co-lex order on $ G /_{\le_G} $ of width $ q $. Moreover, $ \le^\sim_G $ is the maximum co-lex order on $ G /_{\le_G} $ by Corollary \ref{cor:quotientgraphmaximumcolexorder}, and so it is a co-lex order of minimum width. Hence, just build the data structure from Theorem \ref{theor:partialorderspatternmatching} starting from $ G /_{\le_G} $ and $ \le^\sim_G $. Corollary \ref{cor:patternmatchingonquotientisequivalent} ensures that querying $ G /_{\le_G} $ is equivalent to querying $ G $. \qed
\end{proof}

\begin{figure}
	\centering
	\begin{tikzpicture}[shorten >=1pt,node distance=1.6cm,on grid,auto]
	\tikzstyle{every state}=[fill={rgb:black,1;white,10}]
	
	\node[state] (v_1) at (0, 0) {$ v_1 $};
	\node[state] (v_2) at (3, 0) {$ v_2 $};
	\node[rectangle] (v_i) at (6, 0) {$ \dots $};
	\node[state] (v_p) at (9, 0) {$ v_n $};
	\node[state] (u_1) at (0, -3) {$ u_1 $};
	\node[state] (u_2) at (9, -3) {$ u_2 $};
	\path[->]
	(u_1) edge node {a}    (v_1)
	(u_1) edge node {a}    (v_2)
	(u_1) edge node {a}    (v_p)
	(u_2) edge node {a}    (v_1)
	(u_2) edge node {a}    (v_2)
	(u_2) edge node {a}    (v_p)
	;
	\end{tikzpicture}
	\caption{A graph $ G $ for which $ q = 1 $ and $ p = n $. First, we have $ \le_G = \{(v_i, v_j) | 1 \le i, j \le n \} \cup \{(u_1, v_i) | 1 \le i \le n \} \cup \{(u_2, v_i) | 1 \le i \le n \} \cup \{(u_i, u_j) | 1 \le i, j \le 2 \}  $, so $ q = 1 $. Second, let us prove that $ p \ge n $. Let $ \le $ be any co-lex order on $ G $. We must prove that the width of $ \le $ is at least $ n $. Notice that for every $ 1 \le i < j \le n $ nodes $ v_i $ and $ v_j $ are not $ \le $-comparable, because Axiom 2 would imply both $ u_1 < u_2 $ and $ u_2 < u_1 $, which contradicts antisymmetry. Hence $ \{v_1, \dots, v_n \} $ is a $ \le $-antichain and by Dilworth's theorem we conclude that the width of $ \le $ is at least $ n $. The width of $ G $ is indeed $ n $: a co-lex order of width $ n$ is $ \{(u_1, v_i) | 1 \le i \le n \} \cup \{(u_2, v_i) | 1 \le i \le n \} \cup \{(v_i, v_i) | 1 \le i \le n \} \cup \{(u_i, u_i) | 1 \le i \le 2 \} $.}\label{fig:relationbetweenpandq}
\end{figure}
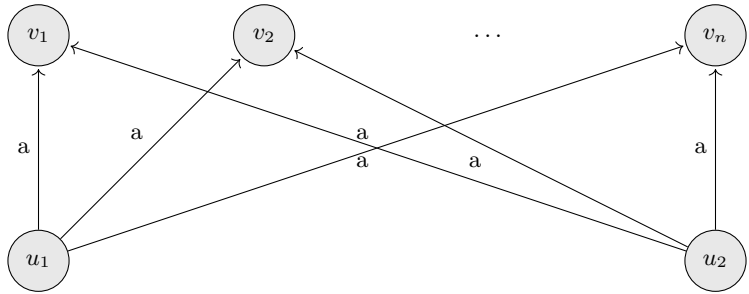

Our Theorem \ref{theor:patternmatchingpreorders} improves on Theorem \ref{theor:partialorderspatternmatching} in several respects:
\begin{enumerate}
    \item For the first time, we show how to build for an \emph{arbitrary} graph a succinct, efficient index for pattern matching \emph{in polynomial time}.
    \item The parameter $ q $ in Theorem \ref{theor:patternmatchingpreorders} is always smaller than or equal to the parameter $ p $ in Theorem \ref{theor:partialorderspatternmatching}, because the maximum co-lex relation on $ G $ refines every co-lex order on $ G $. Moreover, $ q $ can be arbitrarily smaller than $ p $: for every integer $ n $ there exists a graph for which $ q = 1 $ and $ p = n $ (see Figure \ref{fig:relationbetweenpandq}).
    \item The bounds in Theorem \ref{theor:patternmatchingpreorders} only depend on the graph $ G /_{\le_G} $, which may be smaller than the graph $ G $. In other words, \emph{$ G /_{\le_G} $ eliminates the unnecessary redundancy in $ G $ to perform pattern matching}.
\end{enumerate}

\begin{remark}\label{rem:patternmatchingcorrespondencequotient}
In fact, Theorem \ref{theor:patternmatchingpreorders} not only decides whether a pattern occurs in a graph, but if a pattern occurs it also returns indicators for all occurrences, that is, the ($ \le_G^\sim $-convex) set of all nodes at the end of some occurrence. The indicators in the quotient graph can be mapped to indicators in the original graph by storing the correspondence between nodes in the original graph and nodes in the quotient graph.
\end{remark}

\section{Generalizations and applications to automata theory}\label{sec:automata}

Notice that Theorem \ref{theor:patternmatchingpreorders} relies on Corollary \ref{cor:patternmatchingonquotientisequivalent} and not on the more general Lemma \ref{lem:intervalsgointointervals}. However, there are situations where we may interested in matching only patterns that start from a given set of nodes (and not from all nodes). For example, if we consider an automaton, we may be interested in matching strings starting from the initial state, so that we can decide whether a strings belongs to the language recognized by the automaton. Hence, let us generalize Theorem \ref{theor:patternmatchingpreorders}.

\begin{theorem}\label{theor:intervalscolexrelations}
Let $ G = (V, E) $ be a graph, and let $ q $ be the the width of $ \le_G $. Then, there exists a data structure of $|E /_{\le_G}|(\lceil \log|\Sigma|\rceil + \lceil\log q\rceil + 2)\cdot (1+o(1)) + |V /_{\le_G}|\cdot (1+o(1))$ bits, which can be built starting from $ G $  in $ O(|E|^2 + |V /_{\le_G}|^{5 / 2 }) $ time, such that, given a pattern $ P \in \Sigma^* $ and a $ \le_G $-convex set $ U $, in $O(|P| \cdot q^2 \cdot \log(q\cdot |\Sigma|)) $ time decides whether there is an occurrence of $ \alpha $ starting from a node in $ U $.
\end{theorem}

\begin{proof}
Just follow the proof of Theorem \ref{theor:patternmatchingpreorders}, but use Theorem \ref{theor:partialordersindexinggeneral} instead of Theorem \ref{theor:partialorderspatternmatching}. \qed
\end{proof}

\begin{remark}\label{rem:intervalscorrespondencesecond}
Like in Remark \ref{rem:patternmatchingcorrespondencequotient}, if we store the correspondence between nodes in the original graph and nodes in the quotient graph, then Theorem \ref{theor:intervalscolexrelations} also returns indicators for all occurrences of $ \alpha $ starting from a node in $ U $.
\end{remark}

If we want to apply Theorem \ref{theor:intervalscolexrelations} to decide whether a string is accepted by an automaton $ G $, (1) we should make sure the $ \{s \} $, where $ s $ is the initial state, is $ \le_G $-convex, and (2) we should interpret $ G /_{\le_G } $ as an automaton equivalent to $ G $, so that a string belongs to the language recognized by $ G $ if and only if it belongs to the language recognized by $ G /_{\le_G } $.

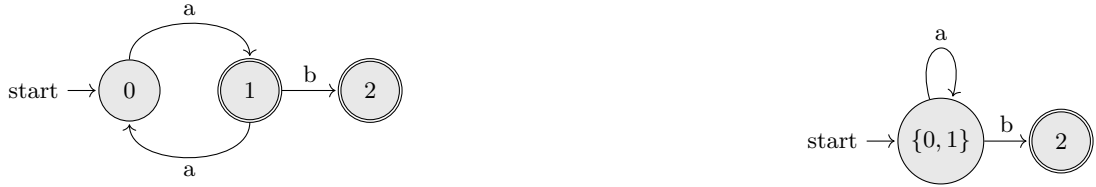
\begin{figure}
	\centering
	     \begin{subfigure}[b]{0.4\textwidth}
         \centering
	\begin{tikzpicture}[shorten >=1pt,node distance=1.6cm,on grid,auto]
	\tikzstyle{every state}=[fill={rgb:black,1;white,10}]
	
	\node[state,initial]   (q_0)                          {$ 0 $};
	\node[state, accepting]           (q_1)  [right of=q_0]    {$ 1 $};
	\node[state, accepting]           (q_2)  [right of=q_1]    {$ 2 $};

	\path[->]
	(q_0) edge [bend left = 90] node {a}    (q_1)
	(q_1) edge [bend left = 90] node {a}    (q_0)
	(q_1) edge node {b}    (q_2);
	\end{tikzpicture}
     \end{subfigure}
     \hfill
     	     \begin{subfigure}[b]{0.4\textwidth}
         \centering
	\begin{tikzpicture}[shorten >=1pt,node distance=1.6cm,on grid,auto]
	\tikzstyle{every state}=[fill={rgb:black,1;white,10}]
	
	\node[state, initial]           (q_1)     {$ \{0, 1 \} $};
	\node[state, accepting]           (q_2)  [right of=q_1]    {$ 2 $};

	\path[->]
	(q_1) edge [loop above] node {a}    (q_1)
	(q_1) edge node {b}    (q_2);
	\end{tikzpicture}
     \end{subfigure}
	\caption{\emph{Left}: an automaton $ G $. \emph{Right}: the automaton $ G /_{\le_G} $. Notice that $ \{0 \} $ is not $ \le_G $-convex because $ [0]_\sim = \{0, 1 \} $ (see Lemma \ref{lem:classesareinconvex}). In fact, if in $ G $ we start from the initial state and read $ b $ we end in $ \emptyset $, whereas in $ G /_\sim $ if we start from the initial state and read $ b $ we end in $ \{ 2 \} $, so the correspondence in Lemma \ref{lem:correspondencequotientcolexrelations} does not hold. Notice also that $ G $ and $ G /_{\le_G} $ do not recognize the same regular language, independently of whether the state $ \{0, 1 \} $ is made final or not.}\label{fig:initialstatenotconvex}
\end{figure}

First, notice that in general $ \{s \} $ is not $ \le_G $-convex, see Figure \ref{fig:initialstatenotconvex} (which also shows that in general $ G /_{\le_G } $ cannot be interpreted as an automaton \emph{equivalent} to $ G $). Let us show that, by adapting the definition of each $ \lambda (v) $, we can force a set to be convex. Let $ G = (V, E) $ be a graph. Let $ @ $ be another special symbol, and assume that $ \# \prec @ \prec a  $ for all $ a \in \Sigma $. Let $ U \subseteq V $ be a subset of nodes, and define:
\begin{equation}\label{eq:newlabelsdefinition}
\resizebox{.93\hsize}{!}{$
    \lambda (v) = 
    \begin{cases}
        \{a \in \Sigma^* | \text{ $ (u, v, a) \in E $ for some $ u \in V $} \} & \text{ if $ v $ has incoming edges and $ v \not \in U $} \\
        \{a \in \Sigma^* | \text{ $ (u, v, a) \in E $ for some $ u \in V $} \} \cup \{@\} & \text{ if $ v $ has incoming edges and $ v \in U $} \\
        \{\# \} & \text{ if $ v $ does not have incoming edges and $ v \not \in U $} \\
        \{\#, @ \} & \text{ if $ v $ does not have incoming edges and $ v \in U $}.
    \end{cases}
$ }
\end{equation}

It is easy to check that all previous results still hold true, with the following remarks:
\begin{enumerate}
    \item The maximum co-lex relation still exists, but in general it is distinct from $ \le_G $. We denote the new maximum co-lex order by $ \le_{G, U} $.
    \item Remark \ref{rem:initialproperty} still holds true, so if $ v \in V $ is such that $ @ \in \lambda (v) $, then it must be $ [v]_\sim = \{v \} $ (because necessarily $ |\lambda (v)| \ge 2 $). When building the graph $ G /_\sim = (V /_\sim, E /_\sim) $ in Definition \ref{def:quotientcolex}, for every $ v \in V $ such that $ @ \in \lambda (v) $ add $ @ $ to $ \lambda ([v]_\sim) $ also. Hence, Remark \ref{rem:propertiescolexquotient} still holds true.
\end{enumerate}

Hence, we can extend Theorem \ref{theor:intervalscolexrelations} as follows.

\begin{theorem}\label{theor:extendedindexingsecondspecialsymbol}
Let $ G = (V, E) $ be a graph, let $ U \subseteq V $, and let $ q $ be the the width of $ \le_{G, U} $. Then, there exists a data structure of $|E /_{\le_{G, U}}|(\lceil \log|\Sigma|\rceil + \lceil\log q\rceil + 2)\cdot (1+o(1)) + |V /_{\le_{G, U}}|\cdot (1+o(1))$ bits, which can be built starting from $ G $  in $ O(|E|^2 + |V /_{\le_{G, U}}|^{5 / 2 }) $ time, such that, given a pattern $ P \in \Sigma^* $ and a $ \le_{G, U} $-convex set $ U $, in $O(|P| \cdot q^2 \cdot \log(q\cdot |\Sigma|)) $ time decides whether there is an occurrence of $ \alpha $ starting from a node in $ U $.
\end{theorem}

Now, it is easy to see that for each $ u \in U $, the set $ \{ u \} $ is $ \le_{G, U} $-convex. Indeed, assume for the sake of contradiction that there existed $ v \in U $ such that $ u < v $ and $ v < u $. In particular, we would have $ [u]_\sim = [v]_\sim $, so by Remark \ref{rem:propertiescolexquotient} we would conclude $ |\lambda (u) | = 1 $, which is a contradiction because $ @ \in \lambda (u) $ implies $ |\lambda (u)| \ge 2 $. Similarly, one can show that $ U $ is $ \le_{G, U} $-convex.

Our next aim is to interpet the quotient automaton as an automaton being equivalent to the original automaton. The quotient automaton should be the quotient graph from Definition \ref{def:quotientcolex} enriched with an initial state and a set of final states.

\begin{definition}\label{def:quotientautomaton}
Let $ \mathcal{A} = (Q, E, s, F) $ be an NFA, and let $ \sim $ be an equivalence relation on $ Q $ such that $ u \sim v $ implies  $ I_u = I_v $. Define $ \mathcal{A /_{\sim}} = (Q /_{\sim}, E /_{\sim}, s /_{\sim}, F /_{\sim}) $ as follows:
\begin{enumerate}
    \item $ Q /_{\sim} = \{[v]_{\sim} | v \in Q \} $;
    \item $ E /_{\sim} = \{([u]_{\sim}, [v]_{\sim}, a )| \text{ $ (u', v', a) \in E $ for some $ u' \in [u]_{\sim} $ and $ v' \in [v]_{\sim}$} \} $;
    \item $ s /_{\sim} = [s]_{\sim} = \{s \} $;
    \item $ F /_{\sim} = \{[v]_\sim \in Q /_{\sim} | \text{ $ v' \in F $ for some $ v' \in [v]_\sim $}   \} $.
\end{enumerate}
\end{definition}

\begin{remark}
Notice that $ [s]_{\sim} = \{s \} $ because there is a string belonging only to $ I_s $, namely, the empy string.
\end{remark}

Let us prove that $ \mathcal{A} $ and $ \mathcal{A /_{\sim}} $ are equivalent. We will also prove a stronger result (that will be useful in the following): $ \mathcal{A} $ and $ \mathcal{A /_{\sim}} $ have the same powerset automaton.

\begin{lemma}\label{lem:relationsbetweenautomatonanditsquotient}
Let $ \mathcal{A} = (Q, E, s, F) $ be an NFA, and let $ \sim $ be an equivalence relation on $ Q $ such that $ u \sim v $ implies  $ I_u = I_q $.
\begin{enumerate}
    \item For clarity, denote by $ I_\alpha $ the set of all states reached by $ \alpha $ on $ \mathcal{A} $ and denote by $ I^\sim_\alpha $ the set of all states reached by $ \alpha $ on $ \mathcal{A /_{\sim}} $. For every $ \alpha \in \Sigma^* $ and for every $ v \in Q $, it holds:
    \begin{equation*}
        v \in I_\alpha \iff [v]_\sim \in I^\sim_\alpha.
    \end{equation*}
    \item For clarity, denote by $ I_v $ the set of all strings that reach $ v $ on $ \mathcal{A} $, and denote by $ I^\sim_{[v]_\sim} $ the set of all strings that reach $ [v]_\sim $ on $ \mathcal{A} $. For every $ v \in V $, it holds:
    \begin{equation*}
        I_v = I^\sim_{[v]_\sim}.
    \end{equation*}
    \item $ \mathcal{L(A /_{\sim})} = \mathcal{L(A)} $.
    \item The powerset automata obtained from $ \mathcal{A /_{\sim}} $ and $ \mathcal{A} $ are isomorphic.
\end{enumerate}
\end{lemma}

\begin{proof}
    \begin{enumerate}
    \item 
    ($ \Rightarrow $) Assume that $ v \in I_\alpha $. We must prove that $ [v]_\sim \in I^\sim_\alpha $. We proceed by induction on $ |\alpha| $. If $ |\alpha| = 0 $, then $ \alpha $ is the empty string $ \epsilon $, so it must be $ v = s $, and indeed $ [s]_\sim \in I^\sim_\epsilon $. Now assume that $ |\alpha| \ge 1 $. We can write $ \alpha = \alpha' a $, with $ \alpha' \in \Sigma^* $ and $ a \in \Sigma $. Since $ v \in I_\alpha $, then there exists $ u \in I_{\alpha'} $ such that $ (u, v, a) \in E $. Hence $ ([u]_\sim, [v]_\sim, a) \in E /_\sim $, and by the inductive hypothesis $ [u]_\sim \in I^\sim_{\alpha'} $, so we conclude $ [v]_\sim \in I^\sim_\alpha $.
    
    ($ \Leftarrow $) Assume that $ [v]_\sim \in I^\sim_\alpha $. We must prove that $ v \in I_\alpha $. We proceed by induction on $ |\alpha| $. If $ |\alpha| = 0 $, then $ \alpha $ is the empty string $ \epsilon $, so it must be $ [v]_\sim = \{s \} $, hence $ v = s $ and indeed $ s \in I_\epsilon $. Now assume that $ |\alpha| \ge 1 $. We can write $ \alpha = \alpha' a $, with $ \alpha' \in \Sigma^* $ and $ a \in \Sigma $. Since $ [v]_\sim \in I^\sim_\alpha $, then there exists $ [u]_\sim \in I^\sim_{\alpha'} $ such that $ ([u]_\sim, [v]_\sim, a) \in E /_\sim $. Hence there exist $ u' \in [u]_\sim $ and $ v' \in [v]_\sim $ such that $ (u', v', a) \in E $. Since $ [u']_\sim = [u]_\sim \in I^\sim_{\alpha'} $, by the inductive hypothesis $ u' \in I_{\alpha'} $, so $ v' \in I_\alpha $. Since $ [v']_\sim = [v]_\sim $, the assumption on $ \sim $ implies that $ I_{v'} = I_v $, hence we conclude $ v \in I_\alpha $.
    \item By point 1, for every $ \alpha \in \Sigma^* $ we have:
    \begin{equation*}
        \alpha \in I_v \iff v \in I_\alpha \iff [v]_\sim \in I^\sim_\alpha \iff \alpha \in I^\sim_{[v]_\sim}.
    \end{equation*}
    \item For every $ \alpha \in \Sigma^* $, we have:
    \begin{equation*}
        \alpha \in \mathcal{L(A /_{\sim})} \iff (\exists [u]_\sim \in F /_\sim)([u]_\sim  \in I^\sim_\alpha) \iff (\exists u \in F)(u \in I_\alpha) \iff \alpha \in \mathcal{L(A)}.
    \end{equation*}
    where the second equivalence holds true because ($ \Leftarrow $) if $ u \in F $ is such that $ u \in I_\alpha $, then $ [u]_\sim \in F /_\sim $ and by point 1 $ [u]_\sim  \in I^\sim_\alpha $, and  $ (\Rightarrow ) $ if $ [u]_\sim \in F /_\sim $ is such that $ [u]_\sim  \in I^\sim_\alpha $, then there exists $ u' \in F $ such that $ u' \in [u]_\sim $, so $ [u']_\sim  \in I^\sim_\alpha $ and by point 1 $ u' \in I_\alpha $.
    \item Let $ (\mathcal{A /_{\sim}})^* $ and $ \mathcal{A}^* $ be the powerset automata obtained from $ \mathcal{A /_{\sim}} $ and $ \mathcal{A} $, respectively. We claim that the function $ f: I_\alpha \mapsto I_\alpha^\sim $ is an isomorphism. Notice that $ f $ is well-defined (that is, $ I_\alpha = I_\beta $ implies $ I^\sim_\alpha = I^\sim_\beta $) and it is bijective by point 1, it respects initial states and edges, and it also respects the final states because $ \mathcal{L(A /_{\sim})} = \mathcal{L(A)} $ by point 3. \qed
    \end{enumerate}
\end{proof}

Let $ \mathcal{A} $ be an NFA. In order to check whether a string is accepted by $ \alpha $, we can use the maximum co-lex relation $ \le_{\mathcal{A}, \{s \}} $. In order to interpret the quotient graph as an automaton, it will suffice to show that the equivalence relation $ \sim_{\le_{\mathcal{A}, \{s \}}} $ has the property required by Lemma \ref{lem:relationsbetweenautomatonanditsquotient}, because then $ \mathcal{A} /_{\le_{\mathcal{A}, \{s \}}} $ can be seen as an automaton equivalent to $ \mathcal{A} $. In other words, we have to prove that $ [u]_{\le_{\mathcal{A}, \{s \}}} = [v]_{\le_{\mathcal{A}, \{s \}}} $ implies $ I_u = I_v $.

We will prove a stronger result: any co-lex relation induces an ordering of the $ I_u $'s. Recall that we assume that $ (\Sigma, \preceq) $ is a totally-ordered alphabet. We can extend this order to $ \Sigma^* $ by sorting strings co-lexicographically (recall that a string $ \alpha $ is co-lexicographically smaller than a string $ \beta $ if the reversed string $ \alpha^R $ is lexicographically smaller than the reversed string $ \beta^R $). Hence, $ (\Sigma^* , \preceq) $ is the total order such that for $ \alpha, \beta \in \Sigma^* $ it holds $ \alpha \preceq \beta $ if and only if the string $ \alpha $ is co-lexicographically smaller than or equal to the string $ \beta $. In \cite{alanko2020regular} it was showed that there is a close relationship between co-lex order of strings and Wheeler orders, and such correspondence motivated the term "co-lex order" \cite{cotumaccio2021indexing}. We now better explicit the role of co-lexicographically sorted strings and we show that the same correspondence also holds for co-lex relations.

\begin{definition}
Let $ \mathcal{A} = (Q, E, s, F) $ be an NFA. Let $ \preceq $ be the reflexive relation on $ \{I_u\  |\ u \in Q \} $ such that, for $ I_u \not = I_v $:
\begin{equation*}
    I_u \prec I_v \iff (\forall \alpha \in I_u)(\forall \beta \in I_v)(\{\alpha, \beta\} \not \subseteq I_u \cap I_v \to \alpha \prec \beta).
\end{equation*}
\end{definition}

Let us prove that we have defined a partial order.

\begin{lemma}\label{lem:orderstringsparialorder}
Let $ \mathcal{A} = (Q, E, s, F) $ be an NFA. Then, $ (\{I_u | u \in Q \}, \preceq) $ is a partial order.
\end{lemma}

\begin{proof}
    Let us prove antisymmetry. Assume that $ I_u \prec I_v $. We must prove that $ I_v \not \prec I_u $. In particular, we have $ I_u \not = I_v $, so there exists $ \alpha \in I_u \setminus I_v $ or $ \beta \in I_v \setminus I_u $. Assume that there exists $ \alpha \in I_u \setminus I_v $ (the other case is analogous). It must be $ I_v \not = \emptyset $, so pick $ \beta \in I_v $. We have $ \{\alpha, \beta \} \not \subseteq I_u \cap I_v $, so $ I_u \prec I_v $ implies $ \alpha \prec \beta $. This means that $ I_v \prec I_u $ cannot hold, otherwise it should also be $ \beta \prec \alpha $, a contradiction because $ (\Sigma^* , \preceq) $ is a total order, and so in particular it is antisymmetric.
    
    Let us prove transitivity. Assume that $ I_u \prec I_v $ and $ I_v \prec I_z $. We must prove that $ I_u \prec I_z $. Pick $ \alpha \in I_u $ and $ \gamma \in I_z $ such that $ \{\alpha, \gamma \} \not \subseteq I_u \cap I_z $. We must prove that $ \alpha \prec \gamma $. Assume that $ \alpha \in I_u \setminus I_z $ (the other case, $ \gamma \in I_z \setminus I_u $, is analogous). We distinguish two cases.
    \begin{enumerate}
        \item Assume $ \alpha \in I_v $. Then $ \alpha \in I_v \setminus I_z $, so $ \{\alpha, \gamma \} \not \subseteq I_v \cap I_z $. From $ I_v \prec I_z $ it follows $ \alpha \prec \gamma $.
        \item Assume $ \alpha \not \in I_v $. We distinguish two subcases.
        \begin{enumerate}
            \item Assume $ \gamma \in I_v $. Then $ \alpha \in I_u \setminus I_v $ and $ \gamma \in I_v $, so $ I_u \prec I_v $ implies $ \alpha \prec \gamma $.
            \item Assume $ \gamma \not \in I_v $. Since $ I_v \not = \emptyset $, pick any $ \beta \in I_v $. We have $ \alpha \in I_u \setminus I_v $ and $ \beta \in I_v $, so $ I_u \prec I_v $ implies $ \alpha \prec \beta $. Moreover, we have $ \gamma \in I_z \setminus I_v $ and $ \beta \in I_v $, so $ I_v \prec I_z $ implies $ \beta \prec \gamma $. From $ \alpha \prec \beta $ and $ \beta \prec \gamma $ we conclude $ \alpha \prec \gamma $ because $ (\Sigma^* , \preceq) $ is a total order, and so in particular it is transitive.
        \end{enumerate}
    \end{enumerate}
\qed
\end{proof}

We can now introduce the class of all relations on the set of states that are co-lexicographically monotonic.

\begin{definition}
Let $\mathcal A = (Q, E, s, F) $. We say that a reflexive relation $ R $ on $ Q $ is \emph{co-lexicographically monotonic} if:
\begin{equation*}
    (u, v) \in R \implies I_u \preceq I_v.
\end{equation*}
\end{definition}

\begin{remark}\label{rem:stringsreachedbymutuallycomparablestatesmustbethesame}
If $ R $ is co-lexicographically monotonic and both $ (u, v) \in R $ and $ (v, u) \in R $, then Lemma \ref{lem:orderstringsparialorder} implies that $ I_u = I_v $. In particular, this holds true if $ \le $ is a co-lexicographically motonic preorder and $ [u]_\le = [v]_\le $.
\end{remark}

Notice that Remark \ref{rem:stringsreachedbymutuallycomparablestatesmustbethesame} implies that if we prove that $ \le_{\mathcal{A}, \{s \}} $ is co-lexicographically monotonic, then $ [u]_{\le_{\mathcal{A}, \{s \}}} = [v]_{\le_{\mathcal{A}, \{s \}}} $ implies $ I_u = I_v $, which is what we want to show. Let us state a more general results: every co-lex relation is co-lexicographically monotonic.

\begin{lemma}\label{lem:colexrelationfromstatestostrings}
Let $\mathcal A = (Q, E, s, F) $ be an NFA, and let $ R $ be a co-lex relation on $ \mathcal{A} $ (with $ U = \{s \} $ in Equation \ref{eq:newlabelsdefinition}). Then, $ R $ is co-lexicographically monotonic.
\end{lemma}

\begin{proof}
        Assume that $ (u, v) \in R $. We must prove that $ I_u \preceq I_v $. If $ I_u = I_v $ we are done, so we can assume $ I_u \not = I_v $ (and in particular $ u \not = v $). Let $ \alpha \in I_u $ and $ \beta \in I_v $ such that $ \{\alpha, \beta \} \not \subseteq I_u \cap I_v $. We must prove that $ \alpha \prec \beta $. Let $ \gamma \in \Sigma^* $ be the longest string such that we can write $ \alpha = \alpha' \gamma $ and $ \beta = \beta' \gamma $, for some $ \alpha', \beta' \in Pref (\mathcal{L(A)}) $. If $ \alpha' = \epsilon $ the conclusion follows, so we can assume $ |\alpha' | \ge 1 $.
        
        Write $ \gamma = c_p \dots c_1 $, with $ c_i \in \Sigma $ for $ i \in \{1, \dots, p \} $ ($ p \ge 0 $). Moreover, write $ \alpha' = a_q \dots a_1 $, with $ a_i \in \Sigma $ for $ i \in \{1, \dots, q \} $ ($ q \ge 1 $), and write $ \beta' = b_r \dots b_1 $, with $ b_i \in \Sigma $ for $ i \in \{1, \dots, r \} $ ($ r \ge 0 $).
        
        Assume $ |\gamma| > 0 $. Since $ \alpha \in I_u $ and $ \beta \in I_v $, then there exist $ u_1, v_1 \in Q $ such that $ \alpha' c_p \dots c_2 \in I_{u_1} $, $ \beta' c_p \dots c_2 \in I_{v_1} $, $ (u_1, u, c_1) \in E $ and $ (v_1, v, c_1) \in E $. By Axiom 2, we obtain $ (u_1, v_1) \in R $. Notice that it must be $ u_1 \not = v_1 $, because $ u_1 = v_1 $ would imply $ \{\alpha, \beta \} \subseteq I_u \cap I_v $. By iterating this argument, we conclude that there exist $ u', v' \in Q $ such that $ \alpha' \in I_{u'} $, $ \beta' \in I_{v'} $ and $ (u', v') \in R $. Clearly, the same conclusion holds also if $ |\gamma| = 0 $, that is, $ \gamma = \epsilon $.
        
        Now, it cannot be $ r = 0 $ because this would imply $ v' = s $, and $ (u', s) \in R $ contradicts Axiom 1 (because $ U = \{s \} $, so $ @ \in \lambda (s) $). Hence, it must be $ |\beta| \ge 1 $. By Axiom 1, it must be $ a_1 \preceq b_1 $. At the same time, the definition of $ \gamma $ implies that it cannot be $ a_1 = b_1 $, so we obtain  $ a_1 \prec b_1 $ and we can conclude $ \alpha \prec \beta $.
        \qed
\end{proof}

We now have all the tools for showing how to decide whether a string belongs to the language recognized by a given automaton. Let $ \mathcal{A} = (Q, E, s, F) $ be an NFA, and let $ \mathcal{A} /_{\le_{\mathcal{A}, \{s \}}} = (Q /_{\le_{\mathcal{A}, \{s \}}}, E /_{\le_{\mathcal{A}, \{s \}}}, s /_{\le_{\mathcal{A}, \{s \}}}, F /_{\le_{\mathcal{A}, \{s \}}}) $ be the quotient NFA from Definition \ref{def:quotientautomaton} obtained by means of the equivalence relation $ \sim_{\le_{\mathcal{A}, \{s \}}} $. Notice that, in fact, $ u \sim_{\le_{\mathcal{A}, \{s \}}} v $ implies $ I_u = I_v $: indeed, $ \le_{\mathcal{A}, \{s \}} $ is co-lexicographically monotonic by Lemma \ref{lem:colexrelationfromstatestostrings}, so the conclusion follows from Remark \ref{rem:stringsreachedbymutuallycomparablestatesmustbethesame}. In particular, by Lemma \ref{lem:relationsbetweenautomatonanditsquotient} automata $ \mathcal{A} $ and $ \mathcal{A} /_{\le_{\mathcal{A}, \{s \}}} $ are equivalent.

\begin{theorem}\label{theor:automatacheckingstring}
Let $\mathcal A = (Q, E, s, F) $ be an NFA, and let $ q $ be the the width of $ \le_{\mathcal{A}, \{s \}} $. Then, there exists a data structure of $|E /_{\le_{\mathcal{A}, \{s \}}}|(\lceil \log|\Sigma|\rceil + \lceil\log q\rceil + 2)\cdot (1+o(1)) + 2|Q /_{\le_{\mathcal{A}, \{s \}}}|\cdot (1+o(1))$ bits, which can be built starting from $ G $  in $ O(|E|^2 + |Q /_{\le_{\mathcal{A}, \{s \}}}|^{5 / 2 }) $ time, that decides in $O(|\alpha| \cdot q^2 \cdot \log(q\cdot |\Sigma|)) $ time whether a string $ \alpha $ belongs to $ \mathcal{L(A)} $.
\end{theorem}

\begin{proof}
Compute $ \le_{\mathcal{A}, \{s \}} $ in $ O(|E|^2) $ time (Theorem \ref{theor:computemaximumrelation}). Build the automaton $  \mathcal{A} /_{\le_{\mathcal{A}, \{s \}}} $ by a graph traversal. By the previous discussion, the quotient automaton is obtained by simply enriching the quotient graph with an initial state and a set of final states, so, recalling that the data structure in Theorem \ref{theor:extendedindexingsecondspecialsymbol} is based on the one in Theorem \ref{theor:partialordersindexinggeneral}, which in turn is based on the one in \cite[Thm 4.2]{cotumaccio2021indexing}, we also need a representation of the final states supporting constant-time rank operations, exactly like in \cite[Thm 4.2]{cotumaccio2021indexing} (see the proof of Theorem \ref{theor:partialordersindexinggeneral}). Then, the conclusion follows from Theorem \ref{theor:extendedindexingsecondspecialsymbol}, because by the previous discussion we know that $ \{s \} $ is $ \le_{\mathcal{A}, \{s \}} $-convex, so we can compute $ I^{\le_{\mathcal{A}, \{s \}}}_\alpha $ (on $ \mathcal{A} /_{\le_{\mathcal{A}, \{s \}}} $) in $O(|\alpha| \cdot q^2 \cdot \log(q\cdot |\Sigma|)) $ time and then check if at least one state in $ I^{\le_{\mathcal{A}, \{s \}}}_\alpha $ is final in $ O(q) $ time (see again \cite[Thm 4.2]{cotumaccio2021indexing}). Lemma \ref{lem:relationsbetweenautomatonanditsquotient} ensures that checking whether $ \alpha $ is recognized by $ \mathcal{A} /_{\le_{\mathcal{A}, \{s \}}} $ is equivalent to checking whether $ \alpha $ is recognized by $ \mathcal{A} $, because the two automata are equivalent. \qed
\end{proof}

\begin{remark}
If one stores the correspondence between states in the original automaton and states in the quotient automata, then Theorem \ref{theor:automatacheckingstring} also returns the set of states accepting a given string (see also Remark \ref{rem:intervalscorrespondencesecond}).
\end{remark}

\begin{remark}
Recall that in this paper we assume that in an automaton all states are reachable from the initial state and each state is either final, or it allows to reach a final state. Note that this assumption is not restrictive for the value of the width of $ \le_{\mathcal{A}, \{s \}} $: if $ R $ is a co-lex relation on an automaton that does not respect these assumptions, then its restriction to the automaton obtained by removing states violating the assumptions is a again a co-lex relation (because no edge entering a state that does not violate the assumptions is removed, so the sets $ \lambda (u) $'s do not change), hence the width of $ \le_{\mathcal{A}, \{s \}} $ cannot increase.
\end{remark}.

Let $ \mathcal{A}^* $ be the DFA equivalent to an NFA $ \mathcal{A} $ obtained by the powerset construction. A classical result \cite{moore} states that in general the number of states of $ \mathcal{A}^* $ is exponential in the number of states of $ \mathcal{A} $. In \cite{cotumaccio2021indexing} is was shown that the number of states is, in fact, exponential only in the minimum width $ p $ of a co-lex order on $ \mathcal{A} $. It is then natural to wonder whether there exists a simpler, smaller parameter capturing the nondeterminism of an automaton. As a first step, one may consider the more general family of co-lex relations. But we can obtain more: what really captures nondeterminism is the property of being co-lexicographically monotonic. Every co-lex relation is co-lexicographically monotonic (Lemma \ref{lem:colexrelationfromstatestostrings}), but a co-lexicographically monotonic relation need not be a co-lex relation. Since our parameter will be, as usual, the width of a relation, it is then natural to consider the \emph{finest} co-lexicographically monotonic relation.

\begin{definition}
Let $ \mathcal{A} = (Q, E, s, F) $ be an NFA. Let $ \preceq_\mathcal{A} $ be the reflexive relation on $ Q $ such that:
\begin{equation*}
    u \prec_\mathcal{A} v \iff I_u \preceq I_v.
\end{equation*}
\end{definition}

From Lemma \ref{lem:orderstringsparialorder}, it follows that $ (Q, \preceq_\mathcal{A}) $ is a preorder. We will see that in order to prove that the width of $ \preceq_\mathcal{A} $ captures the nondeterminism of $ \mathcal{A} $, we would like $ \preceq_\mathcal{A} $ to be a \emph{partial} order. To this end, we will define a quotient automaton $ \mathcal{A} /_{\preceq_\mathcal{A}} $ that captures the nondeterminism of $ \mathcal{A} $ (that is, $ \mathcal{A} $ and $ \mathcal{A} /_{\preceq_\mathcal{A}} $ have the same powerset automaton) and such that $ \preceq_{\mathcal{A} /_{\preceq_\mathcal{A}}} $ is a partial order. Additionally, $ \mathcal{A} /_{\preceq_\mathcal{A}} $ has fewer states that $ \mathcal{A} $, which will contribute to better bounding the number of states of the powerset automaton. Conceptually, this is the the same path that we followed when studying pattern matching on graphs: we first showed that every graph is equivalent to a quotient graph from a pattern matching perspective, then we showed that on the quotient graph the maximum co-lex relation is also the maximum co-lex order.

\begin{figure}
	\centering
	\begin{tikzpicture}[shorten >=1pt,node distance=1.6cm,on grid,auto]
	\tikzstyle{every state}=[fill={rgb:black,1;white,10}]
	
	\node[state,initial]   (q_0)                          {$ 0 $};
	\node[state]           (q_1)  [above right of=q_0]    {$ 1 $};
	\node[state]           (q_2)  [below right of=q_0]    {$ 2 $};
	\node[state, accepting]   (q_3) [right of=q_1] {$ 3 $};
	\node[state, accepting]   (q_4) [right of=q_2] {$ 4 $};
	\path[->]
	(q_0) edge node {c}    (q_1)
	(q_0) edge node {b}    (q_2)
	(q_1) edge node {d}    (q_3)
	(q_1) edge [bend left = 10] node {d}    (q_4)	
	(q_2) edge [bend left = 10] node {d}    (q_3)
	(q_2) edge node {d}    (q_4)
	(q_0) edge [bend left = 90] node {a}    (q_3)
	(q_0) edge [bend right = 90] node [label=below:e] {}    (q_4);
	\end{tikzpicture}
	\caption{We have $ 3 \prec_\mathcal{A} 4 $ (because $ I_3 = \{a, bd, cd \} $ and $ I_4 = \{bd, cd, e \} $) but, $ 3 \not <_{\mathcal{A}, \{s\}} 4, $ (because by Axiom 2 it would imply $ 1 <_{\mathcal{A}, \{s\}} 4 $, but $ \lambda (1) = \{ c \} $ and $ \lambda (2) = \{b \} $ so Axiom 1 would be violated).}\label{fig:stringorderisnotcolex}
\end{figure}
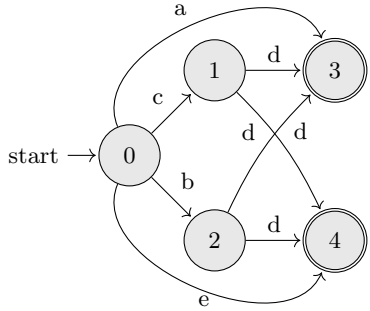

By definition, $ \preceq_\mathcal{A} $ is the finest co-lexicographically monotonic relation on $ \mathcal{A} $. Moreover, from Lemma \ref{lem:colexrelationfromstatestostrings} it follows that $ \preceq_\mathcal{A} $ refines $ \le_{\mathcal{A}, \{s\} } $, and in the general the refinement is strict, see Figure \ref{fig:stringorderisnotcolex}. However, we now prove that if $ \mathcal{A} $ is a DFA, then $ \preceq_\mathcal{A} $ and $ \le_{\mathcal{A}, \{s\} } $ are equal. Note that $ (Q, \preceq_\mathcal{A}) $ is a partial order if $ \mathcal{A} $ is a DFA by Remark \ref{rem:stringsreachedbymutuallycomparablestatesmustbethesame}, because if $ u \not = v $, then $ I_u \not = I_v $ (each $ \alpha \in Pref (\mathcal{L(A)}) $ belongs to exactly one set $ I_u $); in particular, on DFAs $ u \prec_\mathcal{A} v \iff I_u \prec I_v \iff (\forall \alpha \in I_u)(\forall \beta \in I_v)(\alpha \prec \beta) $. Note also that on DFAs Axiom 2 can be equivalently stated as follows: for every $ (u', u, a), (v', v, a) \in E $ such that $ u \not = v $, if $ (u, v) \in R $, then $ u' \not = v' $ and $ (u', v') \in R $.

\begin{lemma}\label{lem:characterizationmaximumcolexrelationDFAs}
Let $ \mathcal{A} = (Q, E, s, F) $ be a DFA. Then, the maximum co-lex relation $ \le_{\mathcal{A}, \{s\}} $ is equal to $ \preceq_\mathcal{A} $, and it is also the maximum colex order.
\end{lemma}

\begin{proof}
Since $ \preceq_\mathcal{A} $ is a partial order on DFAs, it will suffice to prove that $ \preceq_\mathcal{A} $ is the maximum co-lex relation, and since in general $ \preceq_\mathcal{A} $ refines the maximum co-lex relation, it will suffice to prove that $ \preceq_\mathcal{A} $ is a co-lex relation.

Let us prove Axiom 1. Assume that $ u \prec_\mathcal{A} v $. We must prove that $ \lambda (u) \preceq \lambda (v) $. First, notice that it cannot be $ v = s $, because the empty strings belongs to $ I_s $, and it is the co-lexicographically smallest string. This implies that it cannot be $ \# \in \lambda (v) $ or $ @ \in \lambda (v) $. Now, assume that $ a, b \in \Sigma $ are such that $ a \in \lambda (u) $ and $ b \in \lambda (v) $. We must prove that $ a \preceq b $. In particular there exist $ \alpha, \beta \in Pref(\mathcal{L(A)}) $ such that $ \alpha a \in I_u $ and $ \beta b \in I_v $. From $ u \prec_\mathcal{A} v $ we obtain $ \alpha a \prec \beta b $, and so $ a \preceq b $.

Let us prove Axiom 2. Assume that $ (u', u, a), (v', v, a) $ satisfy $ u \prec_\mathcal{A} v $. We must prove that $ u' \prec_\mathcal{A} v' $. Let $ \alpha \in I_{u'} $ and $ \beta \in I_{v'} $. We must prove that $ \alpha \prec \beta $. We have $ \alpha a \in I_u $ and $ \beta a \in I_v $, so from $ u \prec_\mathcal{A} v $ it follows $ \alpha a \prec \beta a $ and so $ \alpha \prec \beta $. \qed
\end{proof}

Let us prove a simple property of co-lexicographically monotonic relations which turns out to be crucial for showing that the width of $ \preceq_\mathcal{A} $ captures the nondeterminism of $ \mathcal{A} $.

\begin{lemma}\label{lem:relationrespectingimpliesconvexsets}
Let $ \mathcal{A} = (Q, E, s, F) $ be an NFA, and let $ R $ be a co-lexicographically monotonic relaation on $ \mathcal{A} $. Then, for every $ \alpha \in Pref (\mathcal{L(A)}) $ the set $ I_\alpha $ is $ R $-convex.
\end{lemma}

\begin{proof}
Assume that $ u, v, z \in Q $ satisfy $ u, z \in I_\alpha $, $ (u, v) \in R $ and $ (v, z) \in R $. We must prove that $ v \in I_\alpha $. Equivalently, we know that $ \alpha \in I_u \cap I_z $, and we must prove that $ \alpha \in I_v $. Since $ R $ is co-lexicographically monotonic, we have $ I_u \preceq I_v $ and $ I_v \preceq I_z $. Suppose by contradiction that $ \alpha \not \in I_v $. In particular, it must be $ I_u \prec I_v \prec I_z $. Pick any $ \beta \in I_v $. We have $ \alpha \in I_u, \beta \in I_v $ and $ \{\alpha, \beta \} \not \subseteq I_u \cap I_v $, so $ I_u \preceq I_v $ implies $ \alpha \prec \beta $. Similarly, we have $ \beta \in I_v $, $ \alpha \in I_z $ and $ \{\beta, \alpha \} \not \subseteq I_v \cap I_z $, so $ I_v \prec I_z $ implies $ \beta \prec \alpha $. This is a contradiction because $ (\Sigma^*, \preceq) $ is a total order, so in particular it is antisymmetric. \qed
\end{proof}

In order to prove that our result on the powerset construction, let us characterize the maximum co-lex relation on a powerset automaton $ \mathcal{A^*}$ (which by Lemma \ref{lem:characterizationmaximumcolexrelationDFAs} is equal to $ \preceq_{\mathcal{A}^*} $).

\begin{lemma}\label{lem:maximalpowerset}
Let $ \mathcal{A} = (Q, E, s, F) $ be an NFA, and let $ \mathcal{A^*} = (Q^*, E^*, s^*, F^*) $ be the powerset automaton obtained from $ \mathcal{A} $. Then, for $ \alpha, \beta \in Pref (\mathcal{L(A)}) $:
\begin{equation*}
    I_\alpha \prec_\mathcal{A^*} I_\beta \iff (\forall \alpha', \beta' \in Pref (\mathcal{L(A)}))((I_{\alpha'} = I_{\alpha}) \land (I_{\beta'} = I_\beta) \to \alpha' \prec \beta')
\end{equation*}
Moreover, for $ \alpha, \beta \in Pref (\mathcal{L(A)}) $ :
\begin{equation*}
    (\exists u \in I_\alpha)(\exists v \in I_\beta)(\{u, v\} \not \subseteq I_\alpha \cap I_\beta \land u \prec_\mathcal{A} v) \implies I_\alpha \prec_\mathcal{A^*} I_\beta.
\end{equation*}
\end{lemma}

\begin{proof}
The first part follows immediately from the definition of $ \preceq_\mathcal{A^*} $ and equation \ref{eq:powersetstates}. Let us prove the second part. Consider $ u \in I_\alpha $ and $ v \in I_\beta $ such that $ \{u, v\} \not \subseteq I_\alpha \cap I_\beta $ and $ u \prec_\mathcal{A} v $. Fix $ \alpha', \beta' \in Pref (\mathcal{L(A)}) $ such that $ I_{\alpha'} = I_\alpha $ and $ I_{\beta'} = I_\beta $. We must prove that $ \alpha' \prec \beta' $. We have $ \alpha' \in I_u $, $ \beta' \in I_v $ and $ \{\alpha', \beta' \} \not \subseteq I_u \cap I_v $, so the conclusion follows from $ u \prec_\mathcal{A} v $. \qed
\end{proof}

Let us define our quotient automaton.

\begin{definition}
Let $ \mathcal{A} = (Q, E, s, F) $ be an NFA. We denote by $ \mathcal{A /_{\preceq_{\mathcal{A}}}} = (Q /_{{\preceq_{\mathcal{A}}}}, E /_{\preceq_{\mathcal{A}}}, s /_{\preceq_{\mathcal{A}}}, F /_{\preceq_{\mathcal{A}}}) $ the NFA from Definition \ref{def:quotientautomaton}, obtained from $ \sim_{\preceq_{\mathcal{A}}} $. We denote by $ \preceq^\sim_\mathcal{A} $ the partial order on $ Q /_{{\preceq_{\mathcal{A}}}} $ induced by the preorder $ \preceq_\mathcal{A} $.
\end{definition}

\begin{remark}
Notice the $ \mathcal{A /_{\preceq_{\mathcal{A}}}} $ is well-defined because because the assumption required in Definition \ref{def:quotientautomaton} holds true for  $ \sim_{\preceq_{\mathcal{A}}} $: if $ [u]_{\preceq_\mathcal{A}} = [v]_{\preceq_\mathcal{A}} $, then $ I_u = I_v $ by Remark \ref{rem:stringsreachedbymutuallycomparablestatesmustbethesame}, being $ \preceq_\mathcal{A} $ a (finest) co-lexicographically monotonic preorder.
\end{remark}

We can now prove that by passing to the quotient automaton the finest co-lexicographically monotic relation becomes a partial order.

\begin{lemma}\label{lem:quotientfinestrespecting}
Let $ \mathcal{A} = (Q, E, s, F) $ be an NFA. Then,  $ \preceq_{\mathcal{A} /_{\preceq_\mathcal{A}}} $ is equal to $ \preceq^\sim_\mathcal{A} $ (so in particular it is a partial order) and it has width equal to the width of $ \preceq_{\mathcal{A}} $.
\end{lemma}

\begin{proof}
By point 2 of Lemma \ref{lem:relationsbetweenautomatonanditsquotient}, for every $ u, v \in Q $ we have:
\begin{equation*}
    [u]_{\preceq_{\mathcal{A}}} \preceq^\sim_\mathcal{A} [v]_{\preceq_{\mathcal{A}}} \iff u \preceq_\mathcal{A} v \iff I_u \preceq I_v \iff I^\sim_{[u]_{\preceq_{\mathcal{A}}}} \preceq I^\sim_{[v]_{\preceq_{\mathcal{A}}}} \iff [u]_{\preceq_{\mathcal{A}}} \preceq_{\mathcal{A} /_{\preceq_\mathcal{A}}} [v]_{\preceq_{\mathcal{A}}}
\end{equation*}
so $ \preceq^\sim_\mathcal{A} $ is equal to $ \preceq_{\mathcal{A /_{\sim_{\mathcal{A}}}}} $. Moreover $ \preceq^\sim_\mathcal{A} $ has the same width of $ \preceq_{\mathcal{A}} $ by Lemma \ref{lem:widthquotientequal}. \qed
\end{proof}

We can finally prove that the number of states of the powerset automaton of $ \mathcal{A} $ is, in fact, exponential in the width of $ \preceq_{\mathcal{A}} $.

\begin{theorem}\label{theor:powersetautomatonexponential in $ r $}
Let $ \mathcal{A} = (Q, E, s, F) $ be an NFA  and let  $ \mathcal{A^*} = (Q^*, E^*, s^*, F^*) $ be the powerset automaton obtained from $ \mathcal{A} $. Let $ n_{\preceq_\mathcal{A}} = |Q /_{{\preceq_{\mathcal{A}}}}| $ and $ n^* = |Q^*| $. Let $ r $ be the width of $ \preceq_\mathcal{A} $ and let $ r^* $ be the width of $ \preceq_\mathcal{A^*} $. Then:
\begin{enumerate}
\item $ r^* \le 2^{r} - 1 $;
\item $ n^* \le 2^{r}(n_{\preceq_\mathcal{A}} - r + 1) - 1 $.
\end{enumerate}
\end{theorem}

\begin{proof}
Since we are interested in obtaining bounds on quantities that refers to the powerset automaton $ \mathcal{A^*} $, by point 4 of Lemma \ref{lem:relationsbetweenautomatonanditsquotient} we can assume without loss of generality that $ \mathcal{A} = \mathcal{A /_{\preceq_{\mathcal{A}}}} $, so $ | Q | = |Q /_{{\preceq_{\mathcal{A}}}}| = n_{\preceq_\mathcal{A}} $ and by Lemma \ref{lem:quotientfinestrespecting} $ \preceq_{\mathcal{A}} $ is a partial order (not only a preorder) and it has width equal to $ r $.

Let $ \{Q_i \}_{i = 1}^{i = r} $ be a $ \preceq_{\mathcal{A}} $-chain partition. For every nonempty $ K \subseteq \{1, \dots, r \} $, define:
\begin{equation*}
    \mathcal{I}_K = \{I_\alpha \in Q^* \mid(\forall i \in \{1, \dots, r \})(I_\alpha \cap Q_i \not = \emptyset \iff i \in K) \}.
\end{equation*}
Notice that $ Q^* $ is the disjoint union of all $ \mathcal{I}_K $. More precisely:
\begin{equation}\label{eq:2expp-1}
    Q^* = \bigsqcup_{\substack{
  \emptyset \not = K \subseteq \{1, \dots, r \}}}
\mathcal{I}_K.
\end{equation}
Let us prove that each $ \mathcal{I}_k $ is a $ \preceq_{\mathcal{A}^*} $-chain. Fix $ I_\alpha, I_\beta \in \mathcal{I}_K $, with $ I_\alpha \not = I_\beta $. We must prove that $ I_\alpha $ and $ I_\beta $ are $ \preceq_{\mathcal{A}^*} $-comparable. Since $ I_\alpha \not = I_\beta $, there exists either $ u \in I_\alpha \setminus I_\beta $ or $ v \in I_\beta \setminus I_\alpha $. Assume that there exists $ u \in I_\alpha \setminus I_\beta $ (the other case is analogous). In particular, let $ i \in \{1, \dots, r \} $ be the unique integer such that $ u \in Q_i $. Since $ I_\alpha, I_\beta \in \mathcal{I}_K $, from the definition of $ \mathcal{I}_K $ it follows that there exists $ v \in I_\beta \cap Q_i $. Notice that $ \{u, v \} \not \subseteq I_\alpha \cap I_\beta $ (so in particular $ u \not = v $), and since $ u, v \in Q_i $ we conclude that $ u $ and $ v $ are $ \preceq_{\mathcal{A}} $-comparable (because $ (Q_i, \preceq_\mathcal{A}) $ is totally ordered, being $ \preceq_\mathcal{A} $ a partial order). By Lemma \ref{lem:maximalpowerset} we conclude that $ I_\alpha $ and $ I_\beta $ are $ \preceq_{\mathcal{A}^*} $-comparable.

\begin{enumerate}
    \item The first part of the theorem follows from equation \ref{eq:2expp-1}, because each $ \mathcal{I}_K $ is a $ \preceq_{\mathcal{A}^*} $-chain and there are $ 2^r - 1 $ choices for $ K $.
    \item Let us prove the second part of the theorem. Fix $ \emptyset \not = K \subseteq \{1, \dots, r \} $. For every $ I_\alpha \in \mathcal{I}_K $ and for every $ i \in K $, let $ m_\alpha^i $ be the smallest element of $ I_\alpha \cap Q_i $ (this makes sense because $ (Q_i, \preceq_\mathcal{A}) $ is totally ordered, being $ \preceq_\mathcal{A} $ a partial order), and let $ M_\alpha^i $ be the largest element of $ I_\alpha \cap Q_i $. Fix $ I_\alpha, I_\beta \in \mathcal{I}_K $, and note the following:
\begin{enumerate}
    \item Assume that for some $ i \in K $ it holds $ m_{\alpha}^i \prec_\mathcal{A} m_{\beta}^i \lor M_{\alpha}^i \prec_\mathcal{A} M_{\beta}^i $. Then, it must be $ I_\alpha \prec_{\mathcal{A}^*} I_\beta $. Indeed, assume that $ m_{\alpha}^i \prec_\mathcal{A} m_{\beta}^i $ (the other case is analogous). We have $ m_{\alpha}^i \in I_{\alpha} $, $ m_{\beta}^i \in I_{\beta} $, $ \{m_{\alpha}^i, m_{\beta}^i \} \not \subseteq I_{\alpha} \cap I_{\beta} $ and $ m_{\alpha'}^i \prec_\mathcal{A} m_{\beta'}^i $, so the conclusion follows from Lemma \ref{lem:maximalpowerset}. Equivalently, we can state that if $ I_\alpha \prec_{\mathcal{A}^*} I_\beta $, then $ (\forall i \in K)(m_{\alpha}^i \preceq_{\mathcal{A}} m_{\beta}^i \land M_{\alpha}^i \preceq_{\mathcal{A}} M_{\beta}^i) $.
    \item Assume that for some $ i \in K $ it holds $ m_{\alpha}^i = m_{\beta}^i \land M_{\alpha}^i = M_{\beta}^i $. Then, it must be $ I_\alpha \cap Q_i = I_\beta \cap Q_i $. Indeed, by Lemma \ref{lem:relationrespectingimpliesconvexsets} the sets $ I_\alpha $ and $ I_\beta $ are $ \preceq_{\mathcal{A}} $-convex (because $ \preceq_{\mathcal{A}} $ is a (finest) co-lexicographically monotonic relation), so $ I_\alpha \cap Q_i $ and $ I_\beta \cap Q_i $ must be equal if their smallest and largest elements on $ (Q_i, \preceq_\mathcal{A}) $ are equal.
    \item Assume that $ (\forall i \in K)(m_{\alpha}^i = m_{\beta}^i \land M_{\alpha}^i = M_{\beta}^i) $. Then, it must be $ I_\alpha = I_\beta $. Indeed, from point (b) we obtain $ (\forall i \in K)(I_\alpha \cap Q_i = I_\beta \cap Q_i) $, so $ I_\alpha = \cup_{i \in K}(I_\alpha \cap Q_i) = \cup_{i \in K}(I_\beta \cap Q_i) = I_\beta $. Notice that we can equivalently state that if $ I_\alpha \not = I_\beta $, then $ (\exists i \in K)(m_{\alpha}^i \not = m_{\beta}^i \lor M_{\alpha}^i \not = M_{\beta}^i) $.
\end{enumerate}
\end{enumerate}

Fix $ I_\alpha, I_\beta \in \mathcal{I}_K $. Now it is easy to show that:
\begin{equation}\label{eq:powersetminmax}
\begin{split}
    I_\alpha \prec_{\mathcal{A}^*} I_\beta & \iff (\forall i \in K)(m_{\alpha}^i \preceq_{\mathcal{A}} m_{\beta}^i \land M_{\alpha}^i \preceq_{\mathcal{A}} M_{\beta}^i) \land \\
    & \land (\exists i \in K)(m_{\alpha}^i \prec_{\mathcal{A}} m_{\beta}^i \lor M_{\alpha}^i \prec_{\mathcal{A}} M_{\beta}^i).
\end{split}
\end{equation}
Indeed, ($ \Leftarrow $) follows from point (a). As for ($ \Rightarrow $), notice that $ (\forall i \in K)(m_\alpha^i \preceq_\mathcal{A} m_{\beta}^i \land M_{\alpha}^i \preceq_\mathcal{A} M_{\beta}^i) $ again follows from point (a), whereas $ (\exists i \in K)(m_{\alpha}^i \prec_\mathcal{A} m_{\beta}^i \lor M_{\alpha}^i \prec_\mathcal{A} M_{\beta}^i) $ follows from point (c).

Let $ |m_{\alpha}^i| $ and $ |M_{\alpha}^i| $ be the positions of $ m_{\alpha}^i $ and $ M_{\alpha}^i $ in the total order $ (Q_i, \preceq_{\mathcal{A}}) $ (so $ |m_{\alpha}^i|, |m_{\alpha}^i| \in \{1, \dots, |Q_i \} $). For every $ I_\alpha \in \mathcal{I}_K $, define:
\begin{equation*}
    T(I_\alpha) = \sum_{i \in K}(|m_\alpha^i| + |M_\alpha^i|).
\end{equation*}
By equation \ref{eq:powersetminmax}, we have that $ I_\alpha \prec_{\mathcal{A}^*} I_\beta $ implies  $ T(I_\alpha) < T(I_\beta) $, so since $ \mathcal{I}_K $ is a $ \preceq_{\mathcal{A}^*} $-chain, we have that $ |\mathcal{I}_K| $ is bounded by the values that $ T(I_\alpha) $ can take. For every $ I_\alpha \in \mathcal{I}_K $ we have $ 2 |K| \le T(I_\alpha) \le 2 \sum_{i \in K} |Q_i| $ (because $ |m_{\alpha}^i|, |m_{\alpha}^i| \in \{1, \dots, |Q_i \} $), so:
\begin{equation}\label{eq:boundIk}
|\mathcal{I}_K| \le 2 \sum_{i \in K} |Q_i| - 2 |K| + 1.
\end{equation}
From equations \ref{eq:2expp-1} and \ref{eq:boundIk}, we obtain:
\begin{equation*}
\begin{split}
    |\mathcal{Q^*}| = \sum_{\emptyset \subsetneqq K \subseteq \{1, \dots, r \}} |\mathcal{I}_K| & \le \sum_{\emptyset \subsetneqq K \subseteq \{1, \dots, r \}} (2 \sum_{i \in K} |Q_i| - 2 |K| + 1) = \\
    & = 2 \sum_{\emptyset \subsetneqq K \subseteq \{1, \dots, r \}} \sum_{i \in K} |Q_i| - 2 \sum_{\emptyset \subsetneqq K \subseteq \{1, \dots, r \}} |K| + \sum_{\emptyset \subsetneqq K \subseteq \{1, \dots, r \}} 1.
\end{split}
\end{equation*}
Notice that $ \sum_{\emptyset \subsetneqq K \subseteq \{1, \dots, r \}} \sum_{i \in K} |Q_i| = 2^{r - 1} \sum_{i \in \{1, \dots, r \}} |Q_i| = 2^{r - 1} n_{\preceq_\mathcal{A}} $ because every $ i \in \{1, \dots, r \} $ occurs in exactly $ 2^{r - 1} $ subsets of $ \{1, \dots, r \} $. Similarly, we obtain $ \sum_{\emptyset \subsetneqq K \subseteq \{1, \dots, r \}} |K| = 2^{r - 1} r $ and $ \sum_{\emptyset \subsetneqq K \subseteq \{1, \dots, r \}} 1 = 2^r - 1 $, We conclude:
\begin{equation*}
    n^* = |\mathcal{Q^*}| \le 2^r n_{\preceq_\mathcal{A}} - 2^r r + 2^r - 1 = 2^r (n_{\preceq_\mathcal{A}} - r + 1) - 1.
\end{equation*}
\qed
\end{proof}

\begin{figure}
	\centering
	\begin{tikzpicture}[shorten >=1pt,node distance=1.6cm,on grid,auto]
	\tikzstyle{every state}=[fill={rgb:black,1;white,10}]
	
	\node[state, accepting] (v_1) at (0, 0) {$ v_1 $};
	\node[state, accepting] (v_2) at (3, 0) {$ v_2 $};
	\node[rectangle] (v_i) at (6, 0) {$ \dots $};
	\node[state, accepting] (v_p) at (9, 0) {$ v_n $};
	\node[state] (u_1) at (0, -3) {$ u_1 $};
	\node[state] (u_2) at (9, -3) {$ u_2 $};
	\node[state, initial] (s) at (4.5, -5) {$ s $};
	\path[->]
	(u_1) edge node {a}    (v_1)
	(u_1) edge node {a}    (v_2)
	(u_1) edge node {a}    (v_p)
	(u_2) edge node {a}    (v_1)
	(u_2) edge node {a}    (v_2)
	(u_2) edge node {a}    (v_p)
	(s) edge node {a}    (u_1)
	(s) edge node {a}    (u_2)
	;
	\end{tikzpicture}
	\caption{One shows that $ p = n $ like in Figure \ref{fig:relationbetweenpandq}. It is immediate to check that $ r = 1 $, because $ I_{s} = \{\epsilon \} $, $ I_{u_1} = I_{u_2} = \{a \} $ and $ I_{v_1} = I_{v_2} = \dots = I_{v_n} = \{aa \} $.} \label{fig:relationbetweenpandr}
\end{figure}
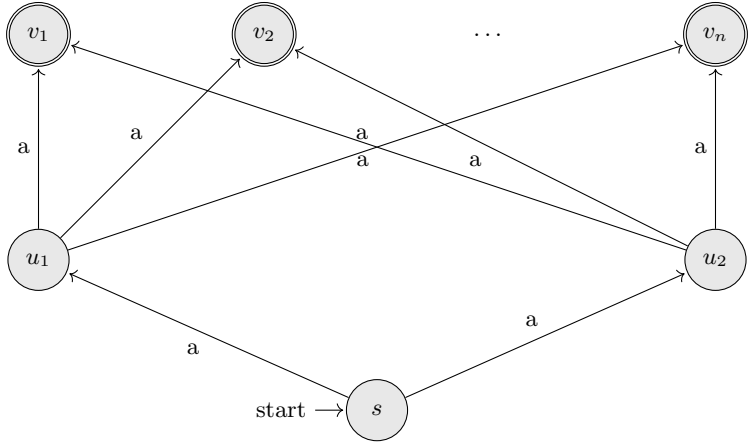

Theorem \ref{theor:powersetautomatonexponential in $ r $} shows that the number of states of $ \mathcal{A^*} $ is exponential not in $ |Q | $, but only in the width $ r $ of $ \preceq_\mathcal{A} $. This result is deeper than the corresponding result in \cite{cotumaccio2021indexing}, where it was showed that the number of states of $ \mathcal{A^*} $ is exponential only in the minimum width $ p $ of a co-lex order on $ \mathcal{A} $: it always holds $ r \le p $ (because $ \preceq_\mathcal{A} $ refines every co-lex order on $ \mathcal{A} $ by Lemma \ref{lem:colexrelationfromstatestostrings}, being $ \preceq_\mathcal{A} $ the finest co-lexicographically monotonic relation on $ \mathcal{A} $), and for every integer $ n $ there exists an NFA such that $ r = 1 $ and $ p = n $ (see Figure \ref{fig:relationbetweenpandr}).

Analogously to what observed in \cite{cotumaccio2021indexing}, one concludes that problems difficult on NFAs but easy on DFAs are fixed-parameter tractable with respect to $ r $. For example, determining whether two NFAs recognize the same language is a PSPACE-complete problem \cite{stockmeyer}. However, the problem is fixed-parameter tractable with respect to $ r $:

\begin{lemma}
Let $ \mathcal{A} = (Q, E, s, F) $ and $ \mathcal{A'} = (Q', E', s', F') $ be NFAs. Let $ n = \max \{|Q|, |Q'| \} $ and $ r = \max \{\text{width of $ \preceq_{\mathcal{A}} $}, \text{width of $ \preceq_{\mathcal{A'}} $} \} $. Then in $O(2^r(n-r+1)n^2 |\Sigma| )$ time we can decide whether $ \mathcal{A} $ and $ \mathcal{A'} $ recognize the same language.
\end{lemma}

\begin{proof}
Follow the proof of \cite[Cor. 5.1]{cotumaccio2021indexing}, and apply the bound in Theorem \ref{theor:powersetautomatonexponential in $ r $}. \qed
\end{proof}

\section{Conclusions and future work}

In this paper we have generalized the idea behind Wheeler graphs to arbitrary edge-labeled graphs. Most importantly, for the first time we have described a polynomial time algorithm that builds a succinct data structure for pattern matching on arbitrary graphs. We have showed that the complexity of the data structure depends on the width of a relation on the set of all nodes. We have also showed that on NFAs the width of a simple relation on the set of states captures the nondeterminism of a given automaton.

Following the seminal paper of Wheeler graphs \cite{GAGIE201767}, all our results assume that we have fixed a total order on the alphabet. However, the pattern matching problem does not require to define such a total order. In general, graphs may be Wheeler with respect to only some orders, and deciding whether a graph is Wheeler with respect to at least one order is a hard problem \cite{dagostino2021ordering}. Even the definition of quotient graph (and quotient automaton) depends on the fixed total order, which suggests that one may obtain smaller and smaller equivalent graphs, all equivalent from a pattern matching perspective, by simply changing the fixed total order and keeping quotienting. On the one hand, in automata theory, an order on the alphabet implies an ordering of the strings accepted by states; on the other hand, from an indexing viewpoint, what really matters is path coherence, which does not require an order on the alphabet. The next natural step is to characterize the class of graphs (broader than the class of Wheeler graphs) that admit an ordering satisfying path coherence. This may also have implications in formal language theory, because Wheeler languages - that is, languages recognized by some Wheeler automaton - are not closed under most boolean operators \cite{alanko2020wheeler}.

Given an NFA $ \mathcal{A} $, we have showed that we can determine $ \le_{\mathcal{A}, \{s\} } $ in polynomial time. However, the relation that best captures the nondeterminism of $ \mathcal{A} $ is $ \preceq_\mathcal{A} $. Computing $ \preceq_\mathcal{A} $ is likely to be a hard problem (because the similar problem of determining whether two automata recognize the same language is a PSPACE-complete). Intuitively, $ \le_{\mathcal{A}, \{s\} } $ describes local properties of states (Axiom 2 only links states and their predecessors), while $ \preceq_\mathcal{A} $ describes global properties of states (the set of all strings accepted by a state depends on the topology of the whole automaton). In order to obtain an estimate of an automaton's nondeterminism, one should try to determine the finest relation lying between $ \le_{\mathcal{A}, \{s\} } $ and $ \preceq_\mathcal{A} $ which is still computable in polynomial time, and bound the width of such a relation as a function of the width of $ \preceq_\mathcal{A} $ (if possible).

\bibliographystyle{plainurl}% the mandatory bibstyle
\bibliography{graphspolynomialindexing.bib}

\end{document}